\documentclass[letterpaper,11pt]{article}
\pdfoutput=1
\usepackage{shorthand}

\usepackage{mathtools}
\usepackage{booktabs}
\usepackage[english]{babel}
\usepackage{amsmath,amssymb,amsbsy,amstext, amsthm, simplewick, amsfonts,braket}
\usepackage{graphicx}
\usepackage[small]{caption}
\usepackage{siunitx}
\usepackage{upgreek}
\usepackage{framed}
\usepackage{wrapfig}
\usepackage{multirow}
\usepackage{bbm}
\usepackage[numbers,sort&compress]{natbib}
\usepackage[svgnames,dvipsnames,x11names]{xcolor}
\usepackage[utf8x]{inputenc}
\usepackage{selinput}
\usepackage{bm}
\usepackage{float}
\usepackage{dsfont}
\usepackage{caption}
\usepackage{subcaption}
\usepackage{sidecap}
\usepackage{longtable}
\usepackage{anyfontsize}

\usepackage{epstopdf}
\usepackage{cancel}
\usepackage{tcolorbox}
\usepackage{latexsym,amsmath,amssymb,epsfig}
\usepackage{braket}
\usepackage{tensor}
\usepackage{tocloft}

\usepackage{tcolorbox}
\definecolor{greyish2}{rgb}{.96,.96,.96}

\def\xyma{\xymatrix@M.7em}
\def\xymas{\xymatrix@M.1em}

\newcommand{\Comment}[1]{{}}
\definecolor{darkblue}{rgb}{0.15,0.35,0.55}
\definecolor{reddish}{rgb}{0.65, 0.2, 0.2}
\definecolor{darkgreen}{RGB}{50,150,0}
\definecolor{greyish2}{rgb}{.96,.96,.96}
\usepackage[linktocpage=true]{hyperref}
\hypersetup{
colorlinks=true,
citecolor=darkblue,
linkcolor=reddish,
urlcolor=darkblue,
pdfauthor={},
pdftitle={},
pdfsubject={}
}

\DeclareFontFamily{OT1}{rsfs10}{}
\DeclareFontShape{OT1}{rsfs10}{m}{n}{ <-> rsfs10 }{}
\DeclareMathAlphabet{\mathscript}{OT1}{rsfs10}{m}{n}

\def\gsim{ \lower .75ex \hbox{$\sim$} \llap{\raise .27ex \hbox{$>$}} }
\def\lsim{ \lower .75ex \hbox{$\sim$} \llap{\raise .27ex \hbox{$<$}} }
\def\be{\begin{equation}}
\def\ee{\end{equation}}
\def\bea{\begin{eqnarray}}
\def\eea{\end{eqnarray}}

\usepackage[a4paper,margin=1in]{geometry}

\usepackage{tikz}
\usetikzlibrary{decorations}
\pgfdeclaredecoration{complete sines}{initial}
{
    \state{initial}[
        width=+0pt,
        next state=upsine,
        persistent precomputation={\pgfmathsetmacro\matchinglength{
            \pgfdecoratedinputsegmentlength / int(\pgfdecoratedinputsegmentlength/\pgfdecorationsegmentlength)}
            \setlength{\pgfdecorationsegmentlength}{\matchinglength pt}
        }] {}
    \state{upsine}[width=\pgfdecorationsegmentlength,next state=downsine]{
        \pgfpathsine{\pgfpoint{0.25\pgfdecorationsegmentlength}{0.5\pgfdecorationsegmentamplitude}}
        \pgfpathcosine{\pgfpoint{0.25\pgfdecorationsegmentlength}{-0.5\pgfdecorationsegmentamplitude}}
    }
    \state{downsine}[width=\pgfdecorationsegmentlength,next state=upsine]{
        \pgfpathsine{\pgfpoint{0.25\pgfdecorationsegmentlength}{-0.5\pgfdecorationsegmentamplitude}}
        \pgfpathcosine{\pgfpoint{0.25\pgfdecorationsegmentlength}{0.5\pgfdecorationsegmentamplitude}}
}
    \state{final}{}
}

\definecolor{greyish}{rgb}{.90,.90,.90}
\definecolor{greyish2}{rgb}{.96,.96,.96}
\usepackage{xcolor,colortbl}
\usepackage{tcolorbox}

\usepackage[all]{xy}


\numberwithin{equation}{section}
\begin{document}
%
%\maketitle
\renewcommand{\thefootnote}{\fnsymbol{footnote}}
~
\vspace{0truecm}
\thispagestyle{empty}

\begin{center}
{\fontsize{20}{24} \bf Exploring $\boldsymbol{2+2}$ Answers to $\boldsymbol{3+1}$ Questions}
\end{center}

\vspace{.15truecm}

\begin{center}
{\fontsize{13}{18}\selectfont
Jonathan J. Heckman,${}^{\rm a}$\footnote{\texttt{\href{mailto:jheckman@sas.upenn.edu}{jheckman@sas.upenn.edu}}}
Austin Joyce,${}^{\rm b}$\footnote{\texttt{\href{mailto:austinjoyce@uchicago.edu}{austinjoyce@uchicago.edu}}}
Jeremy Sakstein,${}^{\rm c}$\footnote{\texttt{\href{mailto:sakstein@hawaii.edu}{sakstein@hawaii.edu}}}
%\\[4.5pt]
and Mark Trodden${}^{\rm a}$\footnote{\texttt{\href{mailto:trodden@physics.upenn.edu}{trodden@physics.upenn.edu}}}
}
\end{center}

\vspace{.2truecm}

 \centerline{{\it ${}^{\rm a}$Center for Particle Cosmology, Department of Physics and Astronomy,}}
 \centerline{{\it University of Pennsylvania, Philadelphia, PA 19104, USA}}

\vspace{.3cm}

   \centerline{{\it ${}^{\rm b}$Kavli Institute for Cosmological Physics, Department of Astronomy and Astrophysics}}
 \centerline{{\it University of Chicago, Chicago, IL 60637, USA} }

\vspace{.3cm}

\centerline{{\it ${}^{\rm c}$Department of Physics and Astronomy,}}
 \centerline{{\it University of Hawai`i, Honolulu, HI 96822, USA} }
 \vspace{.25cm}

\vspace{.3cm}
\begin{abstract}
\noindent
We explore potential uses of physics formulated in Kleinian (i.e., $2+2$) signature spacetimes
as a tool for understanding properties of physics in Lorentzian (i.e., $3+1$) signature. Much as Euclidean (i.e., $4+0$)
signature quantities can be used to formally construct the ground state wavefunction of a Lorentzian signature quantum field theory,
a similar analytic continuation to Kleinian signature constructs a state of low particle flux in the direction of analytic continuation. There is also a natural supersymmetry algebra available in $2+2$ signature, which serves to constrain the structure of correlation functions. Spontaneous breaking of Lorentz symmetry can produce various $\mathcal{N} = 1/2$ supersymmetry algebras that in $3 + 1$ signature correspond to non-supersymmetric systems. We speculate on the possible role of these structures in addressing the cosmological constant problem.

\end{abstract}

\newpage

%\pagenumbering{arabic}
\setcounter{page}{2}
\setcounter{tocdepth}{2}
%\tableofcontents
%\newpage
\renewcommand*{\thefootnote}{\arabic{footnote}}
\setcounter{footnote}{0}

%\section{Introduction}

%\maketitle

%\tableofcontents

\section{Introduction} \label{sec:INTRO}

One of the bedrock principles of physics is that there is just one
time. Indeed, entertaining multiple timelike directions is
tantamount to jeopardizing the whole edifice of causality.\footnote{With
two timelike directions one automatically has
closed timelike curves, leading to numerous pathologies.
Examples include killing your grandfather. Or, being your own
grandfather (but not in the sense of reference \cite{STUPID}). On the other hand, having two timelike directions would make it possible
to bypass various singularities in FLRW cosmology.}
That being said, many formal investigations greatly simplify in Kleinian (i.e.,~$2+2$ signature) spacetimes.
For example, much of the power of modern approaches to scattering amplitudes stems from working with complexified
momenta, and particular simplifications arise in Kleinian signature (see e.g.,~\cite{Penrose:1967wn,Penrose:1968me,Penrose:1985bww, Penrose:1986ca,Parke:1986gb, Dunajski:2001ea, Witten:2003nn,
Arkani-Hamed:2009hub,Monteiro:2020plf,Atanasov:2021oyu,Crawley:2021auj}). Certain string
theories with extended $\mathcal{N}=2$ worldsheet supersymmetry naturally
describe $2+2$ target spacetimes
\cite{Ooguri:1990ww,Ooguri:1991fp,Ooguri:1991ie}, and the
original formulation of F-theory \cite{Vafa:1996xn} (see also
\cite{Castellani:1982ke,Bergshoeff:1982az,Blencowe:1988sk,Bars:1996dz,Bars:1996cm,Hewson:1996yh,
Kutasov:1996zm,Kutasov:1996vh,Tseytlin:1996ne,Bars:1997bz,Bars:1997xb,Nishino:1997gq,Nishino:1997sw,
Hewson:1997wv,Linch:2015lwa,Linch:2015fya,Heckman:2017uxe,Heckman:2018mxl,Heckman:2019dsj})
takes place in an auxiliary $10+2$ signature spacetime. There is also a
precise sense in which $\mathcal{N}=1/2$ supersymmetry can be realized in
Kleinian signature spacetimes~\cite{Heckman:2018mxl,Heckman:2019dsj}, and this
has potential applications to the cosmological constant problem.
Having two times would also provide novel routes to model building, especially in the context of the physics
of extra dimensions \cite{Dvali:1999hn}.

Given this state of affairs, it is fair to ask whether these $2+2$ signature
lessons are simply ``formal tricks,'' or if they have some direct significance in $3+1$
signature. One aim of the present work will be to provide a potential avenue
for connecting $2+2$ physics to our $3+1$ world. Along these lines, we
will explore the sense in which analytic continuation in a spatial direction
is similar to the procedure of continuing to Euclidean signature. In the
latter case, the Euclidean signature path integral can be interpreted as
constructing a wave functional for the ground state of a quantum field theory.
In Kleinian signature, the resulting path integral is not bounded below, but
perturbation theory can still be used to extract the profile of a wave
functional with low particle flux in the direction of analytic continuation.

The other aim of this paper will be to explore some of the structures present
in Kleinian systems. Symmetries of a $2+2$ action serve to constrain
correlation functions via the corresponding Ward identities.~In particular, we study the ways in which
supersymmetry is similar---and also distinct---in this spacetime signature.  Since
the Lorentz group in this space splits as $\mathrm{Spin}(2,2)\simeq \mathrm{SL}(2,\mathbb{R})_{L}\times
\mathrm{SL}(2,\mathbb{R})_{R}$, we can label supersymmetric theories according to the
number of left- and right-handed real spinor generators. The analog of $\mathcal{N}=1$
supersymmetry in Lorentzian signature is therefore instead $\mathcal{N}=(1,1)$
supersymmetry. Spontaneous breaking of the Lorentz group to either chiral
subgroup, $\mathrm{SL}(2,\mathbb{R})_{L}$ or $\mathrm{SL}(2,\mathbb{R})_{R}$, or to the diagonal
subgroup $\mathrm{SL}(2,\mathbb{R})_{D}$ leads to distinct $\mathcal{N}=1/2$
subalgebras.~We comment that the case of chiral $\mathcal{N}=1/2$
supersymmetry is quite similar to the Euclidean signature case
investigated in \cite{Casalbuoni:1975hx, Casalbuoni:1975bj, Casalbuoni:1976tz, Schwarz:1982pf, Ferrara:2000mm,
Klemm:2001yu, Abbaspur:2002xj, deBoer:2003dpn, Ooguri:2003qp,Ooguri:2003tt,Seiberg:2003yz,Britto:2003aj,
Berkovits:2003kj,Cortes:2019mfa,Cortes:2020shr,Gall:2021tiu}.

%Of course,
One of the intriguing features of such $\mathcal{N} = 1/2$ systems
is that in a supersymmetric state, the collection of bubble diagrams corresponding to the quantum corrections to the vacuum energy automatically cancel,
so there is no large cosmological constant problem (at least in Kleinian
signature). The potential use of this, and closely related structures in 3D
$\mathcal{N} = 1$ supersymmetric theories, has been suggested as a way to
protect the ground state from large quantum corrections~\cite{Witten:1994cga, Vafa:1996xn, Heckman:2018mxl, Heckman:2019dsj}.
We present some brief speculative comments on the application of our $2+2$ system to such $3+1$ questions.

\section{$2+2$ and Low Flux States}

In this section we show that the $2+2$ signature path integral constructs a state of low particle
flux of a Lorentzian signature theory. Recall that for a quantum
theory with a bounded Hamiltonian $H$, we can construct the ground state $\left\vert 0\right\rangle$ by acting
on a generic state $\vert \psi \rangle$ in the Hilbert space with the exponentiated Hamiltonian
operator for a long period of time as
\begin{equation}
\left\vert 0\right\rangle \propto \lim_{t\to\infty} e^{-Ht}\left\vert \psi
\right\rangle .
\end{equation}
It is often useful to view the
corresponding vacuum wavefunctional as being constructed by the
path integral:
\begin{equation}
\left\langle \Phi_{f}\left(  \vec{x}\right)  \lvert 0
\right\rangle \sim\int^{\Phi_{f}\left(  \vec{x}\right)
}{\cal D}\phi \, e^{-S_{4,0}[\phi]}, \label{EucPath}%
\end{equation}
where $S_{4,0}$ denotes the action of the Lorentzian theory analytically continued to Euclidean ($4+0$) signature, and where the integration is done over all field configurations that interpolate between the field vanishing in the far past and the field profile $\Phi_f(\vec x)$ at time $t_f$.\footnote{Relatedly, $\lvert \Phi_{f}\left(  \vec{x}\right)\rangle$ denote (Heisenberg picture) field eigenstates, so that $\phi(t_f,\vec x)\lvert \Phi_{f}\left(  \vec{x}\right)\rangle = \Phi_f(\vec x)\lvert \Phi_{f}\left(  \vec{x}\right)\rangle$.}

A similar set of manipulations can be used to construct a class of non-normalizable
wavefunctionals associated with low flux states. We work in a
$(3+1)$-dimensional spacetime with signature $(-,+,+,+)$ and $(2+2)$-dimensional spacetime with signature $(-,+,+,-)$, as obtained
by analytically continuing in the $z$-direction. To illustrate the main idea, consider a single real scalar field in flat
spacetime of $3+1$ signature with the Lagrangian density:
\be
\begin{aligned}
\mathcal{L}_{3,1}  &  =-\frac{1}{2}\eta^{\mu\nu}\partial_{\mu}\phi
\partial_{\nu}\phi-V(\phi)\,,\\
&  =\frac{1}{2}\left(  \partial_{t}\phi\right)  ^{2}-\frac{1}{2}%
\vec{\nabla}\phi\cdot\vec{\nabla}\phi-V(\phi),
\end{aligned}
\ee
where $\vec\nabla$ denotes the spatial gradient. The stress-energy tensor is given by:
\begin{equation}
T_{\mu\nu}=\partial_{\mu}\phi\partial_{\nu}\phi+\eta_{\mu\nu}\mathcal{L}%
_{3,1}.
\end{equation}
In particular, notice that the $tt$ and $zz$ components are the Lagrangian
densities of a scalar field in $4+0$ and $2+2$ signature:\footnote{We comment that some authors prefer a different
sign convention for the Euclidean signature Lagrangian: $\mathcal{L}^{\mathrm{us}}_{4,0} = - \mathcal{L}^{\mathrm{them}}_{4,0}$.
The important physical point is that for either convention, we have a sensible statistical field theory interpretation.}
\begin{align}
T_{tt}  &  =\frac{1}{2}\left(  \left(  \partial_{t}\phi\right)  ^{2}+\left(
\partial_{z}\phi\right)  ^{2}+\left(  \partial_{x}\phi\right)  ^{2}+\left(
\partial_{y}\phi\right)  ^{2}\right)  +V(\phi) \equiv \mathcal{L}_{4,0} \,,
%\equiv \mathcal{L}_{\rm E}
\\
T_{zz}  &  =\frac{1}{2}\left(  \left(  \partial_{t}\phi\right)  ^{2}+\left(
\partial_{z}\phi\right)  ^{2}-\left(  \partial_{x}\phi\right)  ^{2}-\left(
\partial_{y}\phi\right)  ^{2}\right)  -V(\phi) \equiv \mathcal{L}_{2,2} .
\end{align}
Quantizing the $3+1$ signature theory, it is convenient to introduce spatial field configurations
$\Phi\left(  \vec{x}\right)\equiv \phi(\vec x, t_{\ast})  $ at a fixed time $t_{\ast}$.
Similarly, we can represent the conjugate momentum $\pi\left(x\right)
\equiv \partial_{t}\phi(x) |_{t = t_{\ast}}$ as the functional derivative $\Pi(\vec y)= -i\delta/\delta\Phi(\vec y)$.
The field and its momentum then satisfy the canonical commutation relation:
\begin{equation}
[\Phi\left(  \vec{x}\right)  ,\Pi\left(  \vec{y}%
\right)  ]=i\delta^{3}(\vec{x}-\vec{y}).
\end{equation}
Related to $T_{tt}$ and $T_{zz}$, we can write corresponding
operator-valued densities in the field basis:
\begin{align}
\mathcal{H}  &  =\frac{1}{2}\left(  \Pi^{2}+\left(  \partial_{z}\Phi\right)
^{2}+\left(  \partial_{x}\Phi\right)  ^{2}+\left(  \partial_{y}\Phi\right)
^{2}\right)  +V(\Phi)\,,\\
\widetilde{\mathcal{H}}  &  =\frac{1}{2}\left(  \Pi^{2}+\left(  \partial
_{z}\Phi\right)  ^{2}-\left(  \partial_{x}\Phi\right)  ^{2}-\left(
\partial_{y}\Phi\right)  ^{2}\right)  -V(\Phi) \,.
\end{align}
From this, we can also define two integrated operators:
\begin{equation}
H=\int {\rm d}^{3}x\,\mathcal{H}\text{ \ \ and \ \ }\widetilde{H}=\int
{\rm d}^{3}x\,\widetilde{\mathcal{H}}.
\end{equation}
The operator $H$ measures the energy of a state and the operator
$\widetilde{H}$ measures the flux through the $z$ direction in a given state. (Note that $\tl H$ is distinct from
a generator of translations, since these would come
from integrating $T_{0\mu}$ over a spatial slice.)

Given a state
$\left\vert \psi_i \right\rangle $ at an initial time $t_{i}$, we can evolve
it forward using the exponentiated operators:
\begin{align}
U  &  =\exp(-iH\Delta t)\,, & W&=\exp(-H\Delta t)\,,\\
\widetilde{U} &  =\exp(+i\widetilde{H}\Delta t\,)\,, & \widetilde{W}&=\exp(-\widetilde{H}\Delta t\,)\,.%
\end{align}
While $U$ and $\widetilde{U}$ are manifestly unitary, the operators $W$ and
$\widetilde{W}$ instead act as projectors (they
also do not preserve the norm). Given an initial state $\left\vert \psi
_{i}\right\rangle $, we can act on it with the time evolution
operator $U$ and, as is well known, the resulting evolved state is captured by
the standard path integral in Lorentzian signature.
Consider next the evolution generated by acting with $\widetilde{U}$. In this case, the expectation value
$\left\langle \Phi_{f}(\vec{x})\right\vert \widetilde{U}
\cdots \widetilde{U}\left\vert \Phi_{i}(\vec{x})\right\rangle $ can
also be obtained by inserting a basis of eigenstates $\left\vert
\Phi(\vec{x})\right\rangle \left\langle \Phi(\vec{x}
)\right\vert $ and $\left\vert \Pi(\vec{x})\right\rangle
\left\langle \Pi(\vec{x})\right\vert $, and one now obtains:\footnote{Although it is customary
to indicate the path integral with ``limits of integration'' as indicated, the functional integral does not obey a fundamental theorem of calculus if we functionally differentiate with respect to these boundary conditions. Rather, the notation serves as a reminder to sum over all field configurations with prescribed boundary conditions at the beginning and end of a given path.}
\be
\begin{aligned}
\left\langle \Phi_{f}(\vec{x})\right\vert \widetilde{U}%
\cdots \widetilde{U}\left\vert \Phi_{i}(\vec{x})\right\rangle  &
=\int_{\Phi_{i}}^{\Phi_{f}}{\cal D}\phi\,{\cal D}\pi\,\exp\left(  i\int {\rm d}t{\rm d}^{3}x\left[\pi\dot{\phi
}+\widetilde{\mathcal{H}}\right]\right)\,, \label{LegendreTransform}\\
&  =\int_{\Phi_{i}}^{\Phi_{f}}{\cal D}\phi \,e^{i\widetilde{S}}\,,
\end{aligned}
\ee
where, after integrating out the canonical momentum using $\pi = -\dot\phi$ we find
\begin{equation}
\widetilde{S}=\int {\rm d}t\,{\rm d}^{3}x\,\left(  \frac{1}{2}\left[
-\left(  \partial_{t}\phi\right)  ^{2}+\left(  \partial_{z}\phi\right)
^{2}-\left(  \partial_{x}\phi\right)  ^{2}-\left(  \partial_{y}\phi\right)
^{2}\right]  -V(\phi)\right)  .
\end{equation}
Namely, we get back the ``standard'' Lagrangian, but where the roles of $z$ and $t$ have traded places: $z$ is now
functioning effectively as a time coordinate.

\begin{figure}[t!]
\begin{center}
\includegraphics[scale = 1.25, trim = {0cm 0.0cm 0cm 0.0cm}]{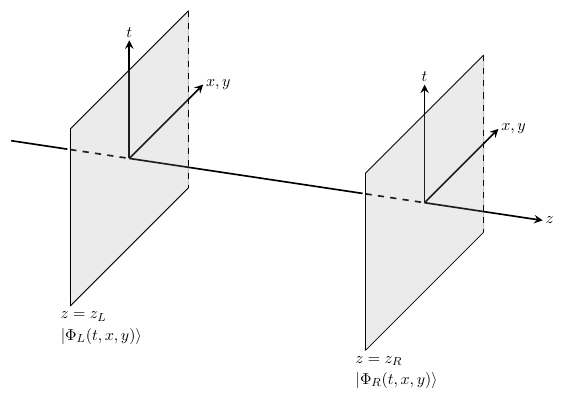}
\caption{Depiction of spatial-slicing of states, indicated at $z = z_L$ and $z = z_R$ respectively as
Schr\"odinger picture states of a $2+1$-dimensional theory $\vert \Phi_{L}(t,x,y)\rangle$ and $\vert \Phi_{R}(t,x,y)\rangle$.}
\label{fig:states}
\end{center}
\end{figure}

In light of the previous discussion,
there is clearly a sense in which acting with $\widetilde{U}$ corresponds to
evolution of a state in the $z$ direction. With this in mind, we now
contemplate a different question: Suppose we slice up our space-time \textit{spatially}
into $2+1$-dimensional systems, indexed by the $z$ direction. We are then free to speak of states (in the $2+1$ Schr\"{o}dinger picture) $\vert \Phi(t,x,y,z_{\ast}) \rangle$, where we fix a reference value of $z_{\ast}$. Consider two such $2+1$ slices separated in the $z$ direction, and labeled as $\vert \Phi_L(t,x,y) \rangle$ and $\vert \Phi_{R}(t,x,y)\rangle$, respectively (see figure \ref{fig:states}).

Now, suppose we are interested in computing expectation values
of field operators which are ordered in the $z$ direction rather than in the
standard time-ordering. That is, given states $\left\vert \Phi_{L}(t,x,y)\right\rangle $
and $\left\vert \Phi_{R}(t,x,y)\right\rangle $ specified at fixed $z$ values,
how would we go about computing:
\begin{equation}
\left\langle \Phi_{L}\left\vert Z\left\{  \phi\left(  x_{1}\right)
\cdots\phi\left(  x_{n}\right)  \right\}  \right\vert \Phi_{R}\right\rangle ,
\end{equation}
where $Z\left\{ \cdots \right\}$ represents $z$-ordering; the $z$ direction equivalent of the usual time-ordering prescription of quantum field theory?

Our proposal is that such quantities can be obtained by evaluating a path
integral expectation value in $2+2$ signature:
\begin{equation}
\left\langle \phi\left(  x_{1}\right)  \cdots \phi\left(  x_{n}\right)
\right\rangle _{L,R}=\frac{\displaystyle\int_{\Phi_{R}}^{\Phi_{L}}{\cal D}\phi\,\exp\left(  -\int
{\rm d}^{4}x\text{ }\mathcal{L}_{2,2}\right)  \phi\left(  x_{1}\right)
\cdots\phi\left(  x_{n}\right)  }{\displaystyle\int_{\Phi_{R}}^{\Phi_{L}}{\cal D}\phi\,\exp\left(
-\int {\rm d}^{4}x\text{ }\mathcal{L}_{2,2}\right)  }.
\end{equation}
We then analytically continue this answer back to $3+1$ signature to obtain
the $z$-ordered correlation function.

We now explain this procedure in more detail. First, we introduce a quantity
closely related to $T_{zz}$ which we shall refer to as a \textquotedblleft
flux\textquotedblright\ functional:
\begin{equation}
\mathcal{F}=\frac{1}{2}\left(  \left(  \partial_{t}\phi\right)  ^{2}%
+\widetilde{\pi}^{2}-\left(  \partial_{x}\phi\right)  ^{2}-\left(
\partial_{y}\phi\right)  ^{2}\right)  -V(\phi),
\end{equation}
Observe that we can now specify a path integral in which the sum over paths
involves boundary conditions $\Phi_{L}(t,x,y)$ and $\Phi_{R}(t,x,y)$
specified at the slices $z = z_L$ and $z_R$. Evolution proceeds
according to this flux $\mathcal{F}$. Indeed, we can introduce the path integral:
\begin{equation}
\int_{\Phi_{R}}^{\Phi_{L}}{\cal D}\phi{\cal D}\tl\pi\,\exp\left(  i\int
{\rm d}^{4}x\Big[\tl\pi\partial_z\phi+\mathcal{F}\Big]\right)  ,
\end{equation}
and integrating out $\widetilde{\pi}$ now results in the standard $3+1$
integrand of the path integral, but where the boundary conditions on the path integral
are now specified at fixed values of the spatial coordinate $z$ rather than at fixed times:
\begin{equation}
\int_{\Phi_{R}}^{\Phi_{L}}{\cal D}\phi\,\exp\left(  i\int {\rm d}^{4}x\text{ }%
\mathcal{L}_{3,1}\right)  .
\end{equation}
Note that whereas in the standard path integral we evolve forward in time,
here we evolve from ``right to left''.\footnote{We would have evolution from left to right
if we had instead evolved with $-\mathcal{F}$.} We can
use this expression to evaluate $z$-ordered (as opposed to time-ordered)
correlation functions of local operators. In this case, the evolution operator
is associated with the \textquotedblleft Hamiltonian density\textquotedblright%
\ $-T_{zz}$, but we emphasize that the corresponding Lagrangian density
appearing in the path integral is identical to the usual case.

We can also compute expectation values with respect to preferred eigenstates
of the flux operator $\mathcal{F}$. Along these lines, we can perform a
path integral:
\begin{equation}
\int_{\Phi_{R}}^{\Phi_{L}}{\cal D}\phi{\cal D}\widetilde{\pi}\,\exp\left(  \int {\rm d}^{4}x\,\Big[i\widetilde{\pi}\partial_{z}
\phi-\mathcal{F}\Big]\right)  =\int_{\Phi_{R}}^{\Phi_{L}}{\cal D}\phi\,\exp\left(  -\int {\rm d}^{4}x\,\mathcal{L}_{2,2}\right)  ,
\end{equation}
where now the Kleinian signature Lagrangian makes an appearance.
\begin{equation}
\mathcal{L}_{2,2}=\frac{1}{2}\left(  \left(  \partial_{t}\phi\right)
^{2}+\left(  \partial_{z}\phi\right)  ^{2}-\left(  \partial_{x}\phi\right)
^{2}-\left(  \partial_{y}\phi\right)  ^{2}\right)  -V(\phi) .
\end{equation}
Since we are slicing up the configuration of paths in a nonstandard way, there is
clearly a sense in which we are making reference to the integrated flux
operator:
\begin{equation}\label{FluxOp}
F(z)=\int {\rm d}t{\rm d}x{\rm d}y \, T_{zz}(t,x,y,z),
\end{equation}
which is something one typically does not do, for the obvious reason that it
grossly violates causality. In particular, integrating over the $t$ direction in equation (\ref{FluxOp})
deviates from the usual formulation of quantum states being
specified on a fixed time slice (i.e., a Cauchy surface). Although it is
clearly a bit formal, a priori there is no issue with treating $F(z)$ as an
operator which acts on our Hilbert space of states. Indeed, since
$T_{zz}(t,x,y,z)$ is constructed from the field $\phi(t,x,y,z)$, and since
$\phi(t,x,y,z)$ is just a linear combination of creation and annihilation
operators, we also get an expression for the operator $F(z)$ in terms of the
same creation and annihilation operators.

Our discussion generalizes to other degrees of freedom. As an illustrative
example, consider a free Dirac field $\psi$. In this case, the relevant
components of the stress-energy tensor are:
\begin{align}
T_{tt}  &  =i\overline{\psi}\gamma_{t}\partial_{t}\psi=-i\overline{\psi
}\left(  \gamma_{z}\partial_{z}+\gamma_{x}\partial_{x}+\gamma_{y}\partial
_{y}\right)  \psi\\
T_{zz}  &  =i\overline{\psi}\gamma_{z}\partial_{z}\psi=-i\overline{\psi
}\left(  \gamma_{t}\partial_{t}+\gamma_{x}\partial_{x}+\gamma_{y}\partial
_{y}\right)  \psi,
\end{align}
where we have used the equations of motion in the second equality. Note that in performing the Legendre transformation
for the flux operator, we introduce a term proportional to $T_{tt}$ in the case of $t$-evolution, and $T_{zz}$ in the case of
$z$-evolution. Combining with the rightmost terms in the above makes manifest that when performing the corresponding evolutions,
the net effect is to analytically continue in the $t$ and $z$ directions, respectively.

The whole discussion can be phrased more abstractly for any field theory
with a stress-energy tensor, $T_{\mu\nu}$. Given a fixed vector $\xi^{\mu}$, we
can evolve our states using the operator $\exp(iF_{\xi
})$ defined via:
\begin{equation}
F_{\xi}=\int {\rm d}^{3}\Sigma^{\mu}\,\xi^{\nu}T_{\mu\nu},
\end{equation}
where $\Sigma^{\mu}$ is the surface perpendicular to $\xi^{\nu}$.
If we instead use the projection operator defined by $\exp(-F_{\xi})$, we see that
successive applications of this operator leads us to states
which have a wave-functional captured by a statistical field theory obtained by
Wick rotation in the $\xi$ direction.\footnote{For example, returning to the scalar field example above, we can work in spherical coordinates and use $T_{rr}$ to project to a state with low flux through constant radius surfaces.}

Clearly, there are three qualitative choices, corresponding to $\xi$ timelike,
spacelike, or null, and the case of present interest to us is the
seemingly most pathological (in the sense of causality), where $\xi$ is spacelike.

When $\xi$ is timelike, $F_{\xi}$ is just the Hamiltonian and this projection operation maps
a generic state onto a linear combination of states with dominant amplitude in the ground state.
What sort of state is being created by instead acting repeatedly with
$\exp(-F_{\xi})$ when $\xi$ is spacelike? To answer this question we decompose the
Hilbert space into eigenstates of $F_{\xi}$. By definition, these
are associated with the pressure or, more precisely, the flux in the $\xi
$ direction. The projection obtained via $\exp(-LF_{\xi})$ for large $L$
amounts to restricting to states with \textquotedblleft
minimal\textquotedblright\ flux in this direction. For all these reasons, we
shall refer to the state obtained by acting with $\exp(-F_{\xi})$ as
\textquotedblleft low flux states in the $\xi$-direction\textquotedblright,
and shall denote them as $\left\vert \text{LOW}\right\rangle $.

Acting with $\exp(-F_{\xi})$ is potentially dangerous when $\xi$ is spacelike because the action
$S_{2,2}$ is unbounded from below. Indeed, the best we can really do is to perform a perturbative analysis around
a given saddle point configuration. The situation is somewhat akin to what occurs in
the Euclidean path integral of gravity, where the action is also unbounded from below.
For some recent additional discussion on some of the subtleties with
analytic continuation of metrics and summing over saddle point configurations,
see e.g., \cite{Kontsevich:2021dmb, Witten:2021nzp}.

Less formally, we expect that actual on-shell physical configurations will always
have a minimal value of $z$-flux. For example, in the case of a perfect fluid
where pressure $p$ is proportional to energy density via $p=w\rho$, the
cosmological constant (which has $w=-1$) would constitute a low pressure configuration.
In order to sidestep these difficulties, we adopt the sentiment that this projection operation is a somewhat formal device that selects low flux states in the vicinity of some chosen saddle point.

Given a $2+2$ Lagrangian, we can also identify candidate symmetries which leave the Lagrangian, and more generally the
partition function, invariant. Such symmetries serve to constrain the correlation functions, and lead to corresponding Ward identities.
The interpretation of these symmetries in $3+1$ signature is more subtle. For example, in a theory with $\mathrm{ISO}(2,2)$ spacetime symmetry, the continuation to Lorentzian signature need not respect this symmetry. That said, the structure of correlation functions will still be constrained.

A related set of issues concerns the systematics of perturbation theory in such systems. Along these lines, suppose we work in $2+2$ signature momentum space. Then, we can present the data of a real four-vector in terms of a complex two-vector with components $(\Omega, K)$. The difference between $2+2$ (Kleinian) and $4+0$ (Euclidean) norms is:
\begin{align}
\mathrm{2+2}: &~~~ k^2 = -\Omega\, \overline{\Omega} + K \overline{K}\\
\mathrm{4+0}: &~~~ k^2 = +\Omega\, \overline{\Omega} + K \overline{K}.
\end{align}
So, much as we can perform all calculations in Lorentzian signature by first Wick rotating to Euclidean signature, we can similarly
perform a Wick rotation from Kleinian to Euclidean signature by formally continuing in the norm $\vert \Omega \vert \rightarrow i \vert \Omega \vert$.

An important subtlety here is the corresponding $i \varepsilon$ prescription when continuing back to Lorentzian signature.~The appropriate notion of $z$-ordering and projection onto the $\vert \mathrm{LOW} \rangle$ state means that in our evaluation of
loop integrals we should work in a deformed contour of integration with respect to the $k_z$ direction. Said differently, we simply apply the standard $i \varepsilon$ prescription, but in the $k_z$ rather than $k_0$ momentum coordinate~\cite{Srednyak:2013ylj}.

\section{$2+2$ Signature Lagrangians}

Having motivated the appearance of $2+2$ signature Lagrangians in some physical problems, we now construct some examples.~We view the action principle as specifying a statistical field theory evaluated around a saddle point
configuration.

Let us begin with the Lagrangian density of a massless free real scalar
field:
\begin{equation}
\mathcal{L}[\phi]=-\frac{1}{2}\eta_{(K)}^{\mu\nu}\partial_{\mu}\phi\partial_{\nu}%
\phi=\frac{1}{2}\left(  \left(  \partial_{t}\phi\right)  ^{2}+\left(
\partial_{z}\phi\right)  ^{2}-\left(  \partial_{x}\phi\right)  ^{2}-\left(
\partial_{y}\phi\right)  ^{2}\right)  ,
\end{equation}
where $\eta_{(K)}^{\mu\nu} = {\rm diag}(-1,+1,+1,-1)$.
By inspection, this is invariant under the Kleinian analog of the Poincar\'e
symmetries, i.e., translations and $\mathrm{Spin}(2,2)$ rotations. Next consider
including additional real scalar fields. We index the fields as
$\phi^{I}$ and introduce a general symmetric constant matrix $M_{IJ}$ for the kinetic terms of
these fields. We have, in general:
\begin{equation}
\mathcal{L}[\phi^{I}]=-\frac{1}{2}M_{IJ}\eta_{(K)}^{\mu\nu}\partial_{\mu}\phi^{I}%
\partial_{\nu}\phi^{J}.
\end{equation}
Since we are working in Kleinian signature, we can a priori allow a
wider variety of possible $M$'s than we would permit in Lorentzian
signature. For example, we could take $M$ to be a $2\times2$ matrix such as:
\begin{equation}
M=\left[
\begin{array}
[c]{cc}%
0 & 1\\
1 & 0
\end{array}
\right]  ,
\end{equation}
which would lead to a Lagrangian with a \textquotedblleft wrong sign kinetic
term\textquotedblright\ for one of our fields. Indeed, starting from such an
action, we can write:
\begin{equation}
\mathcal{L}[\phi^{1},\phi^{2}]=-\eta^{\mu\nu}_{(K)}\partial_\mu\phi^{1}\partial_\nu\phi^{2}=-\frac{1}{2}\eta_{(K)}^{\mu\nu}\partial_\mu
\phi_{+}\partial_\nu\phi_{+}+\frac{1}{2}\eta_{(K)}^{\mu\nu}\partial_\mu\phi_{-}\partial_\nu\phi_{-},
\end{equation}
where we have diagonalized the Lagrangian by  defining the combinations
\begin{equation}
\phi_{\pm}=\frac{1}{\sqrt{2}}\left(  \phi^{1}\pm\phi^{2}\right)  .
\end{equation}
This is of course a pathological Lagrangian in Lorentzian signature (but see also \cite{Carroll:2003st, Cline:2003gs}),
though it will appear quite naturally as the kinetic term of scalars in certain supersymmetric systems in
Kleinian signature.

A priori, in Kleinian signature one could entertain both real and unitary forms of various gauge
groups (as well as suitable complexifications). To see why both possibilities
are available, recall our Lagrangian for a pair of real scalars:
\begin{equation}
\mathcal{L}[\phi^{1},\phi^{2}]=-\eta^{\mu\nu}_{(K)}\partial_\mu\phi^{1}\partial_\nu\phi^{2}\,.
\end{equation}
There is a symmetry of the theory given by the real rescaling:
\begin{equation}
\phi^{1}\mapsto e^{\xi}\phi^{1}\text{ \ \ and \ \ }\phi^{2}\mapsto e^{-\xi
}\phi^{2}.
\end{equation}
The corresponding group of symmetries $\mathbb{R}^{\ast}$ is noncompact. More
generally, we can entertain real forms of various symmetry groups such as
$\mathrm{SL}(N,\mathbb{R})$. There is a sense in which this choice is more natural,
especially in the context of gluon scattering amplitudes in Kleinian signature~\cite{Witten:2003nn,Arkani-Hamed:2009hub}.

Consider next fermionic degrees of freedom. To accomplish
this, we need to discuss spinor representations, associated with the
Clifford algebra:
\begin{equation}
\big\{ \gamma_{(K)}^{\mu},\gamma_{(K)}^{\nu} \big\}  =-2\eta_{(K)}^{\mu\nu
}. \label{gammaKlein}%
\end{equation}
Spinors transform in representations of
$\mathrm{Spin}(2,2)\simeq \mathrm{SL}(2,\mathbb{R})_{L}\times  \mathrm{SL}(2,\mathbb{R})_{R}$. Irreducible
representations are given by real doublets under one of these factors, i.e.,
Majorana--Weyl spinors. Raising and lowering of a spinor index is accomplished with the anti-symmetric
tensors $\varepsilon_{ab}$ and $\varepsilon_{\dot{a} \dot {b}}$ with
$\varepsilon_{12} = -1$.\footnote{For spinors of the $2+1$ signature Lorentz group $\mathrm{SL}(2,\mathbb{R})$,
it is customary to include an additional factor of $i$ in the $\varepsilon_{ab}$ (i.e., charge conjugation)
tensors, with suitable reality conditions enforced via the 3D Dirac matrices. For our purposes, however, where
we still have a notion of chirality (which is not an issue in 3D)
this would be a bit awkward since a doublet $\chi_a$ with
real entries would then be related to the doublet $\chi^a$
with purely imaginary entries.}

Complexifying a Majorana--Weyl spinor results in a Weyl spinor.
Taking a pair of left-handed and right-handed Weyl
spinors gives a Dirac spinor. This is a four-component vector with
complex entries:
\begin{equation}
\Psi=\left[
\begin{array}
[c]{c}%
\psi_{a}\\
\widetilde{\chi}^{\dot{a}}%
\end{array}
\right]  .
\end{equation}
We introduce a Dirac spinor $\Lambda$ in a conjugate representation with
entries:\footnote{Note that because of the choice of signature, we refrain
from introducing the conjugate spinor via $\Psi^{\dag}\cdot\gamma_{(K)}^{0}$.}
\begin{equation}
\Lambda=\left[
\begin{array}
[c]{cc}%
\zeta^{a} & \widetilde{\eta}_{\dot{a}}%
\end{array}
\right]  .
\end{equation}
In terms of this, the Dirac Lagrangian (in any signature, see e.g., reference \cite{Wetterich:2010ni}) is given by:\footnote{A comment on the factor of $-i$. A common practice in the case of the Euclidean signature Dirac action is to absorb the factor of $i$ into $\Lambda$, which is often denoted as ``$\overline{\Psi}$'' even though it is not related to the degrees of freedom in $\Psi$. Our choice to retain the factor of $i$ has to do with subsequent comparison with the literature (for example \cite{Seiberg:2003yz}). Moreover, with the factor of $i$, we can impose a suitable reality condition on the Majorana--Weyl spinor action, something we can achieve in $2+2$ signature, but not $4+0$ signature. For some additional discussion and review of various approaches to Euclidean spinors and Wick rotations, see reference \cite{vanNieuwenhuizen:1996tv}.}
\begin{equation}
\mathcal{L}[\Psi,\Lambda]=-i\Lambda\gamma^{\mu}\partial_{\mu}\Psi.
\end{equation}
By construction, this is Poincar\'{e} invariant. Now we specialize to write down the action of smaller representations. To proceed, it is
helpful to introduce a chiral basis of gamma matrices with all real entries:
\begin{equation}
\gamma_{(K)}^{\mu}=\left[
\begin{array}
[c]{cc}
0& \sigma_{(K)}^{\mu}\\
\overline{\sigma}_{(K)}^{\mu} &0
\end{array}
\right]  ,
\end{equation}
where:
\be
\sigma_{(K)}^{0}   =\left[
\begin{array}
[c]{cc}%
-1 &0 \\
0& -1
\end{array}
\right]  \text{, }~\sigma_{(K)}^{1}=\left[
\begin{array}
[c]{cc}
0& 1\\
1 &0
\end{array}
\right]  \text{, }~\sigma_{(K)}^{2}=\left[
\begin{array}
[c]{cc}%
1 &0 \\
0& -1
\end{array}
\right]  \text{, }~\sigma_{(K)}^{3}=\left[
\begin{array}
[c]{cc}
0& -1\\
1 &0
\end{array}
\right] \,,
\label{eq:paulimatrices}
\ee
and where $\bar\sigma_{(K)}^\mu = (\sigma_{(K)}^0,-\sigma^i_{(K)})$.
Observe that, in contrast to Lorentzian signature, all the $\sigma_{(K)}^{\mu}$'s have
real entries. Moreover, we have analytically continued in the $z$-direction, with
$\sigma_{(K)}^{3}$ anti-Hermitian.\footnote{In comparing with Lorentzian signature conventions, we observe that
$\sigma^2$ and $\sigma^3$ appear to have switched roles. This is just a choice of labelling scheme, and permuting the coordinates would
amount to working with a metric of signature $(-,+,-,+)$. It makes no material difference to the physical content.}
This is required for the gamma matrix algebra to be equation \eqref{gammaKlein}. The Lagrangian for our Dirac fermion now decomposes as:
\begin{align}
\mathcal{L}[\Psi,\Lambda]  &  =-i\left[
\begin{array}
[c]{cc}%
\zeta^{a} & \widetilde{\eta}_{\dot{a}}%
\end{array}
\right]  \cdot\left[
\begin{array}
[c]{cc}
0& \sigma_{(K)}^{\mu}\partial_{\mu}\\
\overline{\sigma}_{(K)}^{\mu}\partial_{\mu} &0
\end{array}
\right]  \cdot\left[
\begin{array}
[c]{c}%
\psi_{a}\\
\widetilde{\chi}^{\dot{a}}%
\end{array}
\right] \\
&  =-i \widetilde{\eta}_{\dot{a}}\left(  \overline{\sigma}_{(K)}^{\mu
}\partial_{\mu}\right)  ^{\dot{a}a}\psi_{a}-i\zeta^{a}\left(  \sigma_{(K)}%
^{\mu}\partial_{\mu}\right)  _{a\dot{a}}\widetilde{\chi}^{\dot{a}}  .
\end{align}
As expected, the action pairs a right-mover with a left-mover, which are in this case complex Weyl spinors.

Now, a special property of split signature is that we can simultaneously impose the Majorana and Weyl conditions.
We are therefore also free to restrict to the special case of purely
real spinors. In what follows, we shall often work with a single pair of Majorana--Weyl spinors
$\lambda_{a}$ and $\widetilde{\lambda}_{\dot{a}}$ and write the action as:
\begin{equation}
\mathcal{L}[\lambda,\widetilde{\lambda}]=-i\widetilde{\lambda}_{\dot{a}}\left(
\overline{\sigma}_{(K)}^{\mu}\partial_{\mu}\right)  ^{\dot{a}a}\lambda_{a}.
\end{equation}
We stress that in Kleinian signature, there is no relation between
$\lambda$ and $\widetilde{\lambda}$ like there is in Lorentzian signature, where they are related by Hermitian conjugation. Rather,
in Kleinian signature they are simply two different
Majorana--Weyl fermions, one of which is left-handed and one of which is right-handed. Note also that as opposed to the situation in
Euclidean signature, here we can enforce a reality condition for the action.\footnote{Indeed, observe that under complex conjugation we have $(\widetilde{\lambda}_{\dot a} (\overline{\sigma}^{\mu}_{(K)} \partial_{\mu})^{\dot{a} a} \lambda_a)^{\ast} = (\overline{\sigma}^{\mu}_{(K)} \partial_{\mu})^{\dot{a} a} \lambda_{a} \widetilde{\lambda}_{\dot a}
= - \widetilde{\lambda}_{\dot a}(\overline{\sigma}^{\mu}_{(K)} \partial_{\mu})^{\dot{a} a} \lambda_{a}$,
where we have used the fact that in Kleinian signature, the $\overline{\sigma}^{\mu}$'s are real matrices,
and the doublets are also real. Including an overall factor of $-i$ then ensures that the action is real.}

Much as in the case of our theory of scalars, we can generalize to multiple fermions.
Assuming that the  kinetic
term is non-degenerate, we can, without loss of generality, use the fermions $\lambda_{a}^{A}$ and
$\widetilde{\lambda}_{\dot{a}}^{B}$ along with a non-degenerate symmetric matrix
$K_{AB}$ to produce:
\begin{equation}
\mathcal{L}[\lambda,\widetilde{\lambda}]=-iK_{AB}\widetilde{\lambda}_{\dot{a}}%
^{A}\left(  \overline{\sigma}_{(K)}^{\mu}\partial_{\mu}\right)  ^{\dot{a}%
a}\lambda_{a}^{B}.
\end{equation}
As in the case of our scalar action, there is no a priori reason to limit our
quadratic form. In what follows, we will suppress the $(K)$ subscript when the
context is clear.

\subsection{Supersymmetry}

Let us now turn to the structure of supersymmetry in Kleinian signature. In
this subsection we focus on the case of $\mathcal{N}=1$ supersymmetry; namely
we have a left-handed Majorana--Weyl spinor $Q_{a}$ and a right-handed
Majorana--Weyl spinor $\widetilde{Q}_{\dot{a}}$. Our conventions in Lorentzian
signature follow those in \cite{Wess:1992cp}, and those of
\cite{Seiberg:2003yz} in Euclidean signature. Our task will be to develop the
related structures in Kleinian signature (see also
\cite{Castellani:1982ke,Bergshoeff:1982az,Blencowe:1988sk,Bars:1996dz,Dunajski:2001ea,Cortes:2019mfa,Cortes:2020shr,Gall:2021tiu}%
). In Kleinian signature, the supersymmetry algebra is:
\begin{align}
 \{ Q_{a},\widetilde{Q}_{\dot{b}}\}   &  =2\sigma_{a\dot{b}}^{\mu
}P_{\mu}& \lbrack P_{\mu},Q_{a}] &  =[P_\mu,\widetilde{Q}_{\dot{a}}]=0\\
\{ Q_{a},Q_{b}\}   &  =\{ \widetilde{Q}_{\dot{a}%
},\widetilde{Q}_{\dot{b}}\}  =0 & \lbrack P_{\mu},P_{\nu}] &  =0,
\end{align}
where $\sigma_{a\dot{b}}^{\mu}$ are the Pauli matrices in Kleinian
signature~\eqref{eq:paulimatrices} (we suppress the $(K)$ subscript) and $P_\mu = i \partial_\mu$.

We note that in contrast to Lorentzian signature, here the spinors are
independent real doublets. Even though they are not related by conjugation, they are still
linked together. For example, observe that there is a redundancy in our
characterization as captured by the rescaling:
\begin{equation}
Q\mapsto e^{\xi}Q\text{ \ \ and \ \ }\widetilde{Q}\mapsto e^{-\xi
}\widetilde{Q}.
\end{equation}
In a theory which respects this rescaling transformation, we have a
corresponding non-compact R-symmetry group $\mathbb{R}^{\ast}$.

Constructing a supersymmetric Lagrangian is also straightforward, and can be
adapted from the Lorentzian signature treatment. Formally speaking we
construct examples of such Lagrangians by replacing all complex conjugate
fields by their tilded versions. This also includes analytic continuation of symmetry groups from compact to real forms.
We begin by introducing the infinitesimal
parameters $\varepsilon_{a}$ and $\widetilde{\varepsilon}_{\dot{a}}$, and use
these to define a symmetry generator:%
\begin{equation}
\delta=\varepsilon^{a}Q_{a}+\widetilde{\varepsilon}_{\dot{a}}\widetilde{Q}%
^{\dot{a}}.
\end{equation}

As an example of a supersymmetric Lagrangian, consider a real scalar $\phi$, a
Majorana--Weyl spinor $\lambda_{a}$ and a real auxiliary field $F$, as well as
partner fields $\widetilde{\phi}$, $\widetilde{\lambda}_{\dot{a}}$ and
$\widetilde{F}$. Our explicit Lagrangian is:
\begin{equation}\label{eq:LAGfree}
\mathcal{L}=-i\widetilde{\lambda}_{\dot{a}}\left(  \overline{\sigma}^{\mu}\partial_{\mu
}\right)  ^{\dot{a}a}\lambda_{a} - \partial^{\mu}\widetilde{\phi}\partial_{\mu
}\phi+\widetilde{F}F.
\end{equation}
We can explicitly verify that this action is invariant under the following
transformation rules:
\begin{align}
\delta\phi &  =\sqrt{2}\varepsilon^{a}\lambda_{a}\text{, \ \ }
&
\delta
\lambda_{a}&= i\sqrt{2}\left(  \sigma^{\mu}\partial_{\mu}\phi\right)  _{a\dot
{a}}\widetilde{\varepsilon}^{\dot{a}}+\sqrt{2}\varepsilon_{a}F\text{,
\ \ }&
\delta F&= i\sqrt{2}\widetilde{\varepsilon}_{\dot{a}}\left(  \overline
{\sigma}^{\mu}\partial_{\mu}\right)  ^{\dot{a}a}\lambda_{a}\,,\\
\delta\widetilde{\phi}  &  =\sqrt{2}\widetilde{\varepsilon}_{\dot{a}%
}\widetilde{\lambda}^{\dot{a}}\,,&
\delta\widetilde{\lambda}_{\dot{a}%
}&=-i\sqrt{2}\epsilon_{\dot a\dot b}\left(  \overline{\sigma}^{\mu}\partial_{\mu}\widetilde{\phi
}\right)  ^{\dot{b}b}\varepsilon_{b}+\sqrt{2}\widetilde{\varepsilon}_{\dot{a}%
}\widetilde{F}\,,~&
\delta\widetilde{F}&=-i\sqrt{2}\varepsilon
^{a}\left(  \sigma^{\mu}\partial_{\mu}\right)_{a\dot{a}}\widetilde{\lambda
}^{\dot{a}}.
\end{align}

To construct supersymmetric actions, it is convenient to work in terms of superspace. To this end,
we supplement our spacetime by a pair of Majorana--Weyl
spinors $\theta^{a}$ and $\widetilde{\theta}^{\dot{a}}$. Using these superspace coordinates, we
have explicit representatives of left and right anti-derivations:
\begin{align}
\label{eq:suspdef1}
Q_{a}  &  =+i\left(  \frac{\partial}{\partial\theta^{a}} + i\sigma_{a\dot{a}%
}^{\mu}\widetilde{\theta}^{\dot{a}}\partial_{\mu}\right)\,,
&
D_{a}  &  =-i\left(  \frac{\partial}{\partial\theta^{a}} -i\sigma_{a\dot{a}%
}^{\mu}\widetilde{\theta}^{\dot{a}}\partial_{\mu}\right)\,,  \\
\widetilde{Q}_{\dot{a}}  &  =-i\left(  \frac{\partial}{\partial
\widetilde{\theta}^{\dot{a}}}+i\theta^{a}\sigma_{a\dot{a}}^{\mu}\partial_{\mu
}\right) \,,
&
\widetilde{D}_{\dot{a}}  &  =+i\left(  \frac{\partial}{\partial
\widetilde{\theta}^{\dot{a}}}-i\theta^{a}\sigma_{a\dot{a}}^{\mu}\partial_{\mu
}\right)  .
\label{eq:suspdef2}
\end{align}
Each of these generators is invariant under complex conjugation because:\footnote{Note that with our
reality conventions, $\left(  {\rm d}\theta\right)  ^{\ast}=-{\rm d}(\theta^{\ast
})=-{\rm d}\theta$. See e.g.,~DeWitt's book on Supermanifolds~\cite{DeWitt:2012mdz}.~This is a consequence of our convention for complex conjugation, namely $(\alpha\beta
)^{\ast}=\beta^{\ast}\alpha^{\ast}$, so for real Grassmann variables, we have
$(\alpha\beta)^{\ast}=-(\alpha\beta)$.}
\begin{equation}
\left(\frac{\partial}{\partial \theta^a} \right)^{\ast} = - \frac{\partial}{\partial \theta^a},\,\,\,\,\, \left(\frac{\partial}{\partial \widetilde{\theta}^{\dot{a}}} \right)^{\ast} = - \frac{\partial}{\partial \widetilde{\theta}^{\dot{a}}},\,\,\,\,\, \left(\frac{\partial}{\partial x^{\mu}} \right)^{\ast} = \frac{\partial}{\partial x^{\mu}}.
\end{equation}
We caution that this is slightly different from Hermitian conjugation, which is implicitly defined by using integration over superspace to define an inner product. Indeed, the Hermitian conjugates have an additional minus sign. The generators $Q_a,\widetilde{Q}_{\dot a},D_a,\widetilde{D}_{\dot a}$'s and $D$'s are thus anti-Hermitian in our conventions. This tracks with what happens in the case of 3D $\mathcal{N} = 1$ superspace (see e.g., equation (2.2.2a) of reference \cite{Gates:1983nr}). It is also instructive
to compare this with what we would have in Lorentzian
signature. There, we would have no overall factor of $i$ or $-i$ on each
term. Additionally, complex conjugation would send $\partial/\partial\theta$
to $-\partial/\partial\overline{\theta}$.

With the definitions~\eqref{eq:suspdef1} and~\eqref{eq:suspdef2}, we obtain the expected
supersymmetry algebra:
\begin{align}
\big\{Q_{a},\widetilde{Q}_{\dot{a}}\big\}   &  =+2i\sigma_{a\dot{a}}^{\mu}\partial_{\mu}\\
\big\{  D_{a},\widetilde{D}_{\dot{a}}\big\}   &  =-2i\sigma_{a\dot{a}}^{\mu}\partial_{\mu}.
\end{align}

It is also convenient to work in terms of a \textquotedblleft
shifted\textquotedblright\ superspace coordinate which favors one handedness
over the other. Introducing coordinates:
\begin{equation}
y^{\mu}=x^{\mu}+i\theta^{a} \sigma^{\mu}_{a\dot{a}}\widetilde{\theta}^{\dot{a}},
\end{equation}
we observe that $y$ is purely real. One can also introduce an ``anti-chiral'':
\begin{equation}
\widetilde{y}^{\mu}=x^{\mu} - i\theta^{a} \sigma^{\mu}_{a\dot{a}}\widetilde{\theta}^{\dot{a}},
\end{equation}
which is also real. In terms of this, the super-derivations
take the form:
\begin{align}
Q_{a}  &  =i\left(  \frac{\partial}{\partial\theta^{a}}\right)
&
D_{a}  &  =i\left(  \frac{\partial}{\partial\theta^{a}}\right) \\
\widetilde{Q}_{\dot{a}}  &  =-i\left(  \frac{\partial}{\partial
\widetilde{\theta}^{\dot{a}}}+2i\theta^{a}\sigma_{a\dot{a}}^{\mu}%
\frac{\partial}{\partial y^{\mu}}\right)
&
\widetilde{D}_{\dot{a}}  &  =-i\left(  \frac{\partial}{\partial
\widetilde{\theta}^{\dot{a}}}-2i\theta^{a}\sigma_{a\dot{a}}^{\mu}%
\frac{\partial}{\partial y^{\mu}}\right)  .
\end{align}

Superfields can be introduced in the standard way. Scalar functions of the superspace coordinates $f(y,\widetilde{y}, \theta, \widetilde{\theta})$ can be expanded into components. For example, chiral and anti-chiral superfields satisfy the conditions:
\begin{align}
\text{Chiral}\text{: } \quad &  \widetilde{D}_{\dot{a}}\Phi=0\,,\\
\text{Anti-Chiral}\text{: } \quad &  D_{a}\widetilde{\Phi}=0\,,
\end{align}
So $\Phi$ depends only on $y$ and $\theta$, and $\widetilde{\Phi}$ depends only on $\widetilde{y}$ and $\widetilde{\theta}$.
In terms of component field expansions, we have:
\begin{equation}
\Phi(y,\theta) = \phi(y) - i \sqrt{2} \theta \lambda(y) - i \theta \theta F(y),
\end{equation}
with a similar expansion for $\widetilde{\Phi}$.

Indeed, in passing from Lorentzian to Kleinian signature, the main difference is that $\Phi$ and $\widetilde{\Phi}$ are not related by complex conjugation. Instead, they are purely real: $\Phi = \Phi^{\dag}$. In this sense, they are more analogous to the superfields of a Lorentzian signature 3D $\mathcal{N} = 1$ system.\footnote{Compared with standard 3D $\mathcal{N} = 1$ conventions (see e.g., \cite{Gates:1983nr}),
we have some additional factors of $i$ in our component field expansion. This is because our $\varepsilon_{ab}$ tensor is purely real.}

To build a supersymmetric action, we can simply make the substitution $\widetilde{\Phi}$ for each complex conjugate quantity $\Phi^{\dag}$ in Lorentzian signature. For example, the Lagrangian for a free superfield $\Phi \oplus \widetilde{\Phi}$ is:
\begin{equation}
\mathcal{L}_{\Phi} = \widetilde{\Phi} \Phi |_{\theta \theta \widetilde{\theta} \widetilde{\theta}},
\end{equation}
where the subscript is an instruction to only keep the term in the component field expansion proportional to $\theta \theta \widetilde{\theta} \widetilde{\theta}$. This yields the Lagrangian of equation (\ref{eq:LAGfree}).

Similar considerations hold for vector superfields, which we specify by the condition $V = V^{\dag}$.
For ease of exposition we focus on the case of abelian gauge fields, the extension to non-abelian gauge symmetry presents no additional
complications. A comment here is that the gaugino is a four-component Majorana fermion,
and in $2+2$ signature, it is more natural to take the real form of the gauge group.\footnote{For example $\mathbb{R}^{\ast}$ rather than $\mathrm{U}(1)$, and $\mathrm{SL}(N,\mathbb{R})$ rather than $\mathrm{SU}(N)$.}
Introducing an abelian vector superfield $V$, gauge transformations act as:
\begin{equation}
V \mapsto V + \Xi - \widetilde{\Xi}
\end{equation}
with $\Xi$ a chiral superfield and $\widetilde{\Xi}$ its counterpart. We can build the spinorial field strengths:
\begin{align}
W_a &= -\frac{1}{4} \widetilde{D} \widetilde{D} D_{a} V \\
\widetilde{W}_{\dot{a}} &= -\frac{1}{4} D D \widetilde{D}_{a} V,
\end{align}
and then construct a corresponding kinetic term via:
\begin{equation}
\mathcal{L}_{\mathcal{W}} = \frac{1}{4 g^2} \left( W^2|_{\theta \theta} + \widetilde{W}^2 |_{\widetilde{\theta} \widetilde{\theta}} \right),
\end{equation}
with $g$ the gauge coupling. We can also couple the vector multiplet to charge $q$ chiral superfields via the term $\widetilde{\Phi} e^{q V} \Phi$. Observe that whereas in Lorentzian and Euclidean signature the gauge transformation on a charged field would send $\Phi \mapsto \exp(- i q \Xi) \Phi$ and $\Phi^{\dag} \mapsto \exp(i q \Xi^{\dag})$ for $\Xi$ a chiral superfield, in Kleinian signature, reality of the various superfields requires us to instead take the transformations $\Phi \mapsto \exp(- q \Xi) \Phi$ and $\widetilde{\Phi} \mapsto \exp(q \widetilde{\Xi}) \widetilde{\Phi}$. Clearly, one can build more elaborate supersymmetric actions for quantum field theories and supergravity theories in the standard way, following, for example reference \cite{Wess:1992cp}. A final comment is that a number of $\mathcal{N} = 2$ actions in various signatures were recently constructed in \cite{Cortes:2019mfa,Cortes:2020shr,Gall:2021tiu}.

\section{Lorentz Breaking and $\mathcal{N}=1/2$ Supersymmetry}

Precisely because the Lorentz group in $2+2$ signature admits spinor
representations with two real degrees of freedom, it is natural to ask whether we can
find Lagrangians which preserve the minimal amount of supersymmetry.
Upon analytic continuation to Lorentzian
signature, we expect such theories to formally have $\mathcal{N}=0$
supersymmetry, but in which some states such as $\left\vert \text{LOW}%
\right\rangle $ are protected against radiative corrections.
This is one of our main observations.

Now, even though the minimal irreducible spinor representation consists of two
real degrees of freedom, building a $\mathrm{Spin}(2,2)$ invariant action in which
degrees of freedom propagate in all four spacetime directions is not actually
possible~\cite{Cortes:2019mfa,Cortes:2020shr, Gall:2021tiu}. This can also be explicitly checked simply by attempting to
construct a kinetic term for a single Majorana--Weyl spinor in Kleinian signature.

To build examples of $\mathcal{N}=1/2$ supersymmetric theories, we must
therefore entertain the possibility of breaking Lorentz symmetry. In fact, the
analogous question in Euclidean signature was studied in  \cite{Ooguri:2003qp, Ooguri:2003tt, Seiberg:2003yz, Berkovits:2003kj}
for a specific breaking pattern based on a self-dual field strength of the
sort which naturally appears in Euclidean signature $\mathcal{N}=2$
supergravity backgrounds with a non-zero graviphoton field strength. In
general terms, there can be breaking patterns triggered by a vector, $v_{a\dot b}$, a
self-dual field strength, $c_{ab}$, or an anti-self-dual field strength, $\tl c_{\dot a\dot b}$, which take the following forms:
\begin{align}
&  \mathrm{Spin}(2,2)\xrightarrow{v_{a\dot{b}}}\mathrm{SL}(2,\mathbb{R})_{D}\,,\\
&  \mathrm{Spin}(2,2)\xrightarrow{c_{ab}}\mathrm{SL}(2,\mathbb{R})_{R}\,,\\
&  \mathrm{Spin}(2,2)\xrightarrow{\widetilde{c}_{\dot{a}\dot{b}}}\mathrm{SL}(2,\mathbb{R})_{L}\,,
\end{align}
these respectively break the\ Lorentz symmetry to either a diagonal or
anti-chiral/chiral subgroup of the full Lorentz group. It is worth noting that there is a sense in which the breaking via a timelike vector is more natural in the context of cosmology. Let us now discuss each possibility in turn.

\subsection{Vector-Like Breaking}

Let us begin with vector-like breaking of the Lorentz group.
Observe that the decomposition of the vector $v_{a\dot{b}}$ under the
diagonal subgroup $\mathrm{SL}(2,\mathbb{R})_{D}$ yields
$\mathbf{3}$~\raisebox{0.145ex}{$\oplus$}~$\mathbf{1}$, so switching on this expectation
value means we need to project to representations of the diagonal group:
\begin{align}
\nonumber
\mathrm{SL}(2,\mathbb{R})_{L}\times \mathrm{SL}(2,\mathbb{R})_{R} &\supset \mathrm{SL}(2,\mathbb{R})_{D} \\
v_{a\dot{b}}:(\mathbf{2},\mathbf{2}) &\rightarrow\mathbf{3}~\displaystyle{\raisebox{0.14ex}{$\oplus$}}~\mathbf{1}\\
Q_{a}:(\mathbf{2},\mathbf{1}) &\rightarrow\mathbf{2}\\
\widetilde{Q}_{\dot{b}}:(\mathbf{1},\mathbf{2}) &\rightarrow\mathbf{2}\,,%
\end{align}
in accord with our general discussion presented above. The presence of $v_{a\dot b}$ as an object in the theory allows us to convert dotted and undotted indices into each other. We should therefore also expect a modified supersymmetry algebra.
Without loss of generality, we may adopt a frame in which
$v_{a\dot{b}} = - \sigma_{a\dot{b}}^{0}$ is
just the identity matrix. Using this tensor, we
can then convert right-handed spinors to left-handed ones via:
\begin{equation}
\overline{Q}_{a} \equiv v_{a\dot{b}}\widetilde{Q}^{\dot{b}}. \label{Qidentification}%
\end{equation}
In other words, we are making the identifications:
\begin{equation}
\overline{Q}_{1}=\widetilde{Q}^{\dot{1}}=\widetilde{Q}_{\dot{2}}\text{ \ \ and
\ \ }\overline{Q}_{2}=\widetilde{Q}^{\dot{2}}=-\widetilde{Q}_{\dot{1}}.
\end{equation}
We can now take a general linear combination of the spinors $\overline
{Q}_{a}$ and $Q_{a}$ as given by:
\begin{equation}
\mathbb{Q}_{a}\equiv \frac{1}{\sqrt{2}}\left(  \zeta Q_{a}+\zeta^{-1}\overline
{Q}_{a}\right)  \text{,}%
\end{equation}
where $\zeta$ is a non-zero real number. This real parameter determines which $\mathcal{N}=1/2$ subalgebra is preserved.

We now ask which supercharges leave our vector $v_{a\dot{b}}$ invariant. In order to facilitate this, we first introduce a corresponding real vector superfield which contains the component:\footnote{Here we are defining $\bar\theta_a \equiv v_{a\dot b}\tl\theta^{\dot b}$. Also, compared with the Lorentzian signature expression we have an additional factor of $-i$ to ensure $V = V^{\dag}$.}
\begin{equation}
V= \cdots + i\theta^{a}v_{a\dot{b}}\widetilde{\theta}^{\dot{b}}+ \cdots = \cdots + i\theta
^{a}\overline{\theta}_{a}+ \cdots\,. \label{Vexpand}%
\end{equation}
Acting with $\mathbb{Q}_{a}$ results in:
\begin{equation}
\mathbb{Q}_{a}\, V=\frac{1}{\sqrt{2}}\left(  \zeta\overline{\theta}%
_{a}-\zeta^{-1}\theta_{a}\right)  .
\end{equation}
So, we see that there is a 3D $\mathcal{N}=1$ superspace invariant under the
action of the $\mathbb{Q}_{a}$'s given by identifying:
\begin{equation}
\zeta\overline{\theta}_{a}=\zeta^{-1}\theta_{a}.
\end{equation}

The end result of this is that the $\mathcal{N}=1/2$ algebra is just that of a
3D $\mathcal{N}=1$ system:
\be
\begin{aligned}
\left\{  \mathbb{Q}_{a},\mathbb{Q}_{b}\right\}   &  =\frac{1}{2}\left\{  \zeta
Q_{a}+\zeta^{-1}\overline{Q}_{a},\zeta Q_{b}+\zeta^{-1}\overline{Q}%
_{b}\right\} \\
&  =\frac{1}{2}\left\{  Q_{a},\overline{Q}_{b}\right\}  +\frac{1}{2}\left\{
\overline{Q}_{a},Q_{b}\right\} \\
&  =v_{b\dot{b}}P_{a}^{~\,\dot{b}}+v_{a\dot{a}}P_{b}^{~\,\dot{a}},
\end{aligned}
\ee
which we summarize by simply writing the 3D $\mathcal{N}=1$ algebra:
\begin{equation}
\left\{  \mathbb{Q}_{a},\mathbb{Q}_{b}\right\}  =2P_{ab},
\end{equation}
where $P_{ab}$ generates translations in the direction transverse to the timelike vector $v_{a \dot{b}}$.

We now discuss a few generalizations. The analysis just presented assumed a timelike vector, but in Kleinian signature we could equally well
have considered a spacelike vector. The reason is that under spatial breaking, we would retain a
3D Lorentz symmetry group with metric of signature $(-,-,+)$. The situation is different in Lorentzian signature.
Activating a timelike vector would leave us with a metric of signature $(+,+,+)$, and spinors in 3D Euclidean space are pseudo--real rather than real. Imposing a reality condition on a pseudo--real doublet would force all entries to be zero. On the other hand, a spacelike vector
can preserve 3D $\mathcal{N} = 1$ supersymmetry, as happens for supersymmetric domain walls.

Vector-like breaking can also be accomplished for any background which produces the same breaking pattern of the Lorentz group.
Indeed, the general form of the coset construction for spontaneously broken spacetime symmetries ensures that other choices, e.g., a time-dependent rolling scalar background, as well as a three-form flux will also result in the same symmetry breaking pattern. A related comment
is that in the F-theory motivated scenario of \cite{Heckman:2018mxl, Heckman:2019dsj}, a three-form flux threads the spatial slice of an
FLRW cosmology, which, upon analytically continuing to Kleinian signature, retains the same $\mathcal{N} = 1/2$ supersymmetry considered here.

\subsubsection{Example}

Let us now give an example of vector-like breaking. To keep the supersymmetry of the system manifest,
we embed our vector field in a non-dynamical vector superfield $V$, and keep it at a fixed background value.\footnote{In other words, we view the real vector $v_{\mu}$ as the field strength for a zero-form potential. In terms of superfields we can write $V = - i \theta D S - i \widetilde{\theta} \widetilde{D} \widetilde{S} $ for a background chiral multiplet $S$ and its partner $\widetilde{S}$. We have included both $S$ and $\widetilde{S}$ in anticipation of continuing back to Lorentzian signature.} We can couple matter fields to $V$ to obtain a corresponding supersymmetric action. To illustrate, consider a chiral multiplet (and its partner) $\Phi \oplus \widetilde{\Phi}$ with respective charges $+q$ and $-q$ under the vector.
The supersymmetric kinetic term for the chiral multiplet is:\footnote{So far we have been following the Lorentzian signature conventions of \cite{Wess:1992cp}, but to avoid various factors of $1/2$ for covariant derivatives in later expressions
we now rescale $V \mapsto -2V$.}
\begin{equation}
S=\int {\rm d}^{4}x{\rm d}^{4}\theta\text{ }\widetilde{\Phi}e^{-2qV}\Phi + \cdots\,.
\end{equation}
Giving $V$ a background value:
\begin{equation}
V= i \theta^{a}v_{a\dot{b}}\widetilde{\theta}^{\dot{b}}= i \theta^{a}\overline{\theta}_{a},
\end{equation}
we note that only two real supercharges will leave this function of superspace
invariant. Indeed, that is the content of our discussion around equation~\eqref{Vexpand}.
Plugging this expression back into our action, we obtain the following component field action:
\begin{equation}
S =\int {\rm d}^{4}x\text{ } -i \widetilde{\lambda}_{\dot{a}}\left(
\partial^{\dot{a}a}+ q v^{\dot{a}a}\right)  \lambda_{\dot{a}} -(\partial_{\mu}- q v_{\mu})\widetilde{\phi}(\partial^{\mu
}+ q v^{\mu})\phi +\cdots\,,
\end{equation}
where we have expanded in the component fields of the various superfields. For example, $\phi$ is the scalar component of
$\Phi$ and $\lambda_a$ is its fermionic superpartner.
Additionally, we have used the bispinor notation $\partial^{\dot{a} a} = \overline{\sigma}_{\mu}^{\dot{a} a} \partial^{\mu}$.
This leads to Lorentz violating couplings, but this is potentially permissible if it is solely in the timelike direction (as occurs anyway in cosmology).

\subsection{Chiral Breaking}

Chiral breaking of the Lorentz group retains an $\mathcal{N}=1/2$
subalgebra simply because one half of the superalgebra generated by $Q$ and $\widetilde{Q}$
is not deformed at all. Euclidean signature effective actions were
constructed using the language of non-anti-commutative Grassmann variables in
superspace in references \cite{Ooguri:2003qp,Ooguri:2003tt,Seiberg:2003yz}. In more
pedestrian terms, we can simply construct the corresponding supersymmetric
action compatible with spontaneous breaking of a chiral half of the Lorentz algebra.
Deformations of the underlying superspace geometry have been considered in a number of references,
including \cite{Casalbuoni:1975hx, Casalbuoni:1975bj, Casalbuoni:1976tz, Schwarz:1982pf, Ferrara:2000mm,
Klemm:2001yu, Abbaspur:2002xj, deBoer:2003dpn, Ooguri:2003qp, Ooguri:2003tt, Kawai:2003yf, Chepelev:2003ga,
David:2003ke, Seiberg:2003yz, Britto:2003aj,Berkovits:2003kj}, although we add that these references
consider Euclidean, rather than Kleinian signature spacetime.

The general approach in this setup is to consider a deformation of the Grassmann superspace coordinates:
\begin{equation}
\{\theta^a , \theta^b \} = c^{ab},
\end{equation}
with $c^{ab}$ a constant background field. This can also be accompanied by a non-commutative deformation of the spacetime coordinates, namely
$[x^{\mu} , x^{\nu}] = i b^{\mu \nu}$. In reference \cite{Seiberg:2003yz} the construction of effective actions in terms of a corresponding generalized Moyal product was developed. From the present perspective, this is but one choice of symmetry breaking for the Lorentz group.

Physical backgrounds which produce this sort of breaking pattern are obtained from switching on a background self-dual field strength. This can be arranged in both Euclidean and Kleinian signature. For further details on the analysis in Euclidean signature, see e.g., \cite{Ooguri:2003qp, Ooguri:2003tt, Kawai:2003yf, Chepelev:2003ga, David:2003ke, Seiberg:2003yz}.

\section{Spatial Instabilities} \label{ssec:Instabilities}

A different application of $2+2$ signature Lagrangians is in the study of
Lorentzian signature systems with a spatial instability.\footnote{We thank
C.L. Kane for discussions on this point.} To give an example, consider a system
of $(2+1)$-dimensional theories arranged as \textquotedblleft coupled
layers\textquotedblright\ in a $(3+1)$-dimensional system with one lattice
direction. For concreteness, we index the layers by $j\in\mathbb{Z}$ so that
for each layer we have fields $\phi_{j}$, and the action is:
\begin{equation}
S=\underset{j}{\sum}S_{j}+\underset{j}{\sum}S_{j,j+1},
\end{equation}
where $S_{j}$ is the action for a free field on a single layer:
\begin{equation}
S_{j}=\int {\rm d}^{3}x\,\frac{1}{2}\left(  \left(  \partial_{t}\phi
_{j}\right)  ^{2}-\left(  \partial_{x}\phi_{j}\right)  ^{2}-\left(
\partial_{y}\phi_{j}\right)  ^{2}\right)  ,
\end{equation}
and $S_{j,j+1}$ is the contribution from nearest neighbor interaction term:
\begin{equation}
S_{j,j+1}=\int {\rm d}^{3}x\left(-\frac{\alpha}{4}\sin^{2}\left(  \phi_{j}%
-\phi_{j+1}\right) \right)  ,
\end{equation}
which we we can treat as a bounded effective potential. When $\alpha>0$, this is
just giving a lattice approximation to a $(3+1)$-signature system, and the
ground state has $\left\langle \phi_{j}-\phi_{j+1}\right\rangle =0$. For
$\alpha<0$, this same configuration is actually a local maximum and the minimum
is instead reached by taking $\phi_{j}-\phi_{j+1}=\pi/2$. Expanding around
this local maximum, we obtain a $2+2$ signature Lagrangian density.

Instanton configurations in $2+2$ signature correspond to transitions from one local maximum to another. This has a clear meaning in the context of the 3D layers construction, but also generalizes to other field theories, including Yang-Mills theory. Indeed,
in both $2+2$ and $4+0$ signature, there are self-dual field configurations.

We remark that this construction is compatible with 3D $\mathcal{N} = 1$ supersymmetry.
Introducing 3D real superfields $\Phi_{j}$, we can couple neighboring layers via the superpotential:
\begin{equation}
W_{\mathrm{eff}} = \underset{j}{\sum}  \mathrm{cos}(\Phi_{j} - \Phi_{j+1}),
\end{equation}
which implements a superspace version of dimensional deconstruction, much as in~\cite{Arkani-Hamed:2001vvu,Dijkgraaf:2003xk}.
The resulting effective potential $V \sim \vert \partial W / \partial \Phi \vert^2$ is positive definite, but when expanded around a local maximum will result in a deconstructed Kleinian signature theory.

\section{Discussion}

One of the original motivations for this work was to better understand the
potential role of $\mathcal{N}=1/2$ supersymmetry in $2+2$ signature
spacetimes as a way to address the \textquotedblleft cosmological constant
problem\textquotedblright. In $2+2$ signature,
bubble diagrams that correct the energy of a $\mathcal{N}=1/2$ supersymmetric state automatically vanish.
Indeed, this fact was already observed in the case of
chiral symmetry breaking of supersymmetry in Euclidean signature in
\cite{Seiberg:2003yz,Britto:2003aj,Terashima:2003ri}, but it clearly extends
to other choices of Lorentz breaking such as those which are of relevance in
cosmological backgrounds.

What might this mean for a Lorentzian signature spacetime? From the
present perspective, a perturbative calculation in $2+2$ signature is related
to correlation functions computed in a specific class of low-flux states. The
statement that $\mathcal{N}=1/2$ supersymmetry is retained in $2+2$ signature
means that these states do not mix at the quantum level with other states.
From this perspective, it is tempting to speculate that rather than working
with the ground state of a quantum field theory, the low flux states are more
appropriate in the context of cosmology.

Of course, one of the important phenomenological implications of supersymmetry
is the prediction of new superpartners for all of the states of the Standard
Model of particle physics. From the way we have constructed our $\mathcal{N}%
=1/2$ Lagrangians in $2+2$ signature, we see that the field content for the
$3+1$ theory obtained from analytic continuation will have precisely the same
degrees of freedom as in the MSSM, with the caveat that we do not expect to
have as much control over the resulting superpartner masses.

There are two conflicting intuitions, which make it challenging to reliably extract
the mass spectrum.~References \cite{Heckman:2018mxl,
Heckman:2019dsj} suggested that the geometric mean of an IR\ and UV\ cutoff
could arise via F-theory on a $\mathrm{Spin}(7)$ background, which in turn could
produce superpartner masses on the order of $10-100$ TeV. Are the
considerations presented here compatible with these coarse expectations?

On the one hand, in Minkowski space there is no such thing as
$\mathcal{N}=1/2$ supersymmetry. From this perspective, it is natural to
suspect that the $\mathcal{N}=1/2$ MSSM just involves a high scale for
supersymmetry breaking, once radiative corrections to the superpartner masses
are taken into account. On the other hand, the explicit models of
$\mathcal{N}=1/2$ supersymmetry we investigated involved spontaneous breaking
of Lorentz symmetry, which, to be compatible with cosmology, must lead to a
rather low scale for supersymmetry breaking. Said differently, in all of our
explicit realizations, we constructed $\mathcal{N}=1$ Lagrangians which only
broke to $\mathcal{N}=1/2$ supersymmetry at very low cosmological scales. Even so,
it is well-known in other contexts that even seemingly mild Lorentz breaking
terms can, after including radiative corrections, produce relatively large effects which are often difficult to reconcile with observational constraints.

One clear indication of an $\mathcal{N}=1/2$ structure would be apparent Lorentz violation effects, which would in turn
suggest an apparent violation of CPT symmetry (see e.g., \cite{Colladay:1996iz}).
It would be interesting to study the phenomenological
consequences of this, and related aspects of an $\mathcal{N}=1/2$ MSSM.

In this paper we have laid out a framework for thinking about quantum field theories in Kleinian signature spacetimes, describing the drawbacks, advantages, and challenges of exploiting these structures to address problems in $3+1$-dimensional physics. In future work we will take on some of these major challenges, constructing and studying an $\mathcal{N}=1/2$ MSSM, understanding the mass spectrum, and further interrogating the formal structure of this approach.

%\newpage

\vspace{-.25cm}
\paragraph{Acknowledgments:}

We thank G. Gabadadze, C.L. Kane,  J. Stout, A. Tomasiello, E. Torres and S. Wong for helpful
discussions. This work was initiated at the 2019 Penn Center for Particle Cosmology retreat
hosted by PDT partners in New York, and we thank the local hosts, especially
D. Wesley for kind hospitality. Some of this work was also performed at the
2022 Aspen winter conference on Geometrization of (S)QFTs in $D\leq6$ held at
the Aspen Center for Physics, which is supported by National Science
Foundation grant PHY-1607611. The work of JJH and MT is supported by DOE (HEP)
Award DE-SC0013528. The work of AJ is supported by DOE (HEP) Award DE-SC0009924.
The work of MT is also supported by the Simons Foundation
Origins of the Universe Initiative, grant number 658904.

\appendix

\newpage

\bibliographystyle{utphys}
\bibliography{TwoTwo}

\providecommand{\href}[2]{#2}\begingroup\raggedright\begin{thebibliography}{10}

\bibitem{STUPID}
T.~Arnold, ``{\textit{The Stupids}: I'm My Own Grandpa},'' {\em Savoy Pictures}
  (1996)  .

\bibitem{Penrose:1967wn}
R.~Penrose, ``{Twistor Algebra},''
  \href{http://dx.doi.org/10.1063/1.1705200}{{\em J. Math. Phys.} {\bf 8}
  (1967)  345}.

\bibitem{Penrose:1968me}
R.~Penrose, ``{Twistor Quantization and Curved Space-Time},''
  \href{http://dx.doi.org/10.1007/BF00668831}{{\em Int. J. Theor. Phys.} {\bf
  1} (1968)  61--99}.

\bibitem{Penrose:1985bww}
R.~Penrose and W.~Rindler,
  \href{http://dx.doi.org/10.1017/CBO9780511564048}{{\em {Spinors and
  Space-Time. Vol. 1}}}.
\newblock Cambridge Monographs on Mathematical Physics. Cambridge Univ. Press,
  Cambridge, UK, 1985.

\bibitem{Penrose:1986ca}
R.~Penrose and W.~Rindler,
  \href{http://dx.doi.org/10.1017/CBO9780511524486}{{\em {Spinors and
  Space-Time. Vol. 2}}}.
\newblock Cambridge Monographs on Mathematical Physics. Cambridge University
  Press, Cambridge, UK, 1986.

\bibitem{Parke:1986gb}
S.~J. Parke and T.~R. Taylor, ``{An Amplitude for $n$ Gluon Scattering},''
  \href{http://dx.doi.org/10.1103/PhysRevLett.56.2459}{{\em Phys. Rev. Lett.}
  {\bf 56} (1986)  2459}.

\bibitem{Dunajski:2001ea}
M.~Dunajski, ``{Anti-self-dual four-manifolds with a parallel real spinor},''
  \href{http://dx.doi.org/10.1098/rspa.2001.0918}{{\em Proc. Roy. Soc. Lond. A}
  {\bf 458} (2002)  1205--1222}, \href{http://arxiv.org/abs/math/0102225}{{\tt
  arXiv:math/0102225}}.

\bibitem{Witten:2003nn}
E.~Witten, ``{Perturbative Gauge Theory As A String Theory In Twistor Space},''
  \href{http://dx.doi.org/10.1007/s00220-004-1187-3}{{\em Commun. Math. Phys.}
  {\bf 252} (2004)  189--258}, \href{http://arxiv.org/abs/hep-th/0312171}{{\tt
  arXiv:hep-th/0312171}}.

\bibitem{Arkani-Hamed:2009hub}
N.~Arkani-Hamed, F.~Cachazo, C.~Cheung, and J.~Kaplan, ``{The S-Matrix in
  Twistor Space},'' \href{http://dx.doi.org/10.1007/JHEP03(2010)110}{{\em JHEP}
  {\bf 03} (2010)  110}, \href{http://arxiv.org/abs/0903.2110}{{\tt
  arXiv:0903.2110 [hep-th]}}.

\bibitem{Monteiro:2020plf}
R.~Monteiro, D.~O'Connell, D.~Peinador~Veiga, and M.~Sergola, ``{Classical
  Solutions and their Double Copy in Split Signature},''
  \href{http://dx.doi.org/10.1007/JHEP05(2021)268}{{\em JHEP} {\bf 05} (2021)
  268}, \href{http://arxiv.org/abs/2012.11190}{{\tt arXiv:2012.11190
  [hep-th]}}.

\bibitem{Atanasov:2021oyu}
A.~Atanasov, A.~Ball, W.~Melton, A.-M. Raclariu, and A.~Strominger, ``{(2, 2)
  Scattering and the Celestial Torus},''
  \href{http://dx.doi.org/10.1007/JHEP07(2021)083}{{\em JHEP} {\bf 07} (2021)
  083}, \href{http://arxiv.org/abs/2101.09591}{{\tt arXiv:2101.09591
  [hep-th]}}.

\bibitem{Crawley:2021auj}
E.~Crawley, A.~Guevara, N.~Miller, and A.~Strominger, ``{Black Holes in Klein
  Space},'' \href{http://arxiv.org/abs/2112.03954}{{\tt arXiv:2112.03954
  [hep-th]}}.

\bibitem{Ooguri:1990ww}
H.~Ooguri and C.~Vafa, ``{Selfduality and $\mathcal{N}=2$ String {MAGIC}},''
  \href{http://dx.doi.org/10.1142/S021773239000158X}{{\em Mod. Phys. Lett. A}
  {\bf 5} (1990)  1389--1398}.

\bibitem{Ooguri:1991fp}
H.~Ooguri and C.~Vafa, ``{Geometry of $\mathcal{N}=2$ Strings},''
  \href{http://dx.doi.org/10.1016/0550-3213(91)90270-8}{{\em Nucl. Phys. B}
  {\bf 361} (1991)  469--518}.

\bibitem{Ooguri:1991ie}
H.~Ooguri and C.~Vafa, ``{$\mathcal{N}=2$ Heterotic Strings},''
  \href{http://dx.doi.org/10.1016/0550-3213(91)90042-V}{{\em Nucl. Phys. B}
  {\bf 367} (1991)  83--104}.

\bibitem{Vafa:1996xn}
C.~Vafa, ``{Evidence for F theory},''
  \href{http://dx.doi.org/10.1016/0550-3213(96)00172-1}{{\em Nucl. Phys. B}
  {\bf 469} (1996)  403--418}, \href{http://arxiv.org/abs/hep-th/9602022}{{\tt
  arXiv:hep-th/9602022}}.

\bibitem{Castellani:1982ke}
L.~Castellani, P.~Fre, F.~Giani, K.~Pilch, and P.~van Nieuwenhuizen, ``{Beyond
  $d=11$ Supergravity and Cartan Integrable Systems},''
  \href{http://dx.doi.org/10.1103/PhysRevD.26.1481}{{\em Phys. Rev. D} {\bf 26}
  (1982)  1481}.

\bibitem{Bergshoeff:1982az}
E.~Bergshoeff, M.~de~Roo, and B.~de~Wit, ``{Conformal Supergravity in
  Ten-dimensions},'' \href{http://dx.doi.org/10.1016/0550-3213(83)90159-1}{{\em
  Nucl. Phys. B} {\bf 217} (1983)  489}.

\bibitem{Blencowe:1988sk}
M.~P. Blencowe and M.~J. Duff, ``{Supermembranes and the Signature of
  Space-time},'' \href{http://dx.doi.org/10.1016/0550-3213(88)90155-1}{{\em
  Nucl. Phys. B} {\bf 310} (1988)  387--404}.

\bibitem{Bars:1996dz}
I.~Bars, ``{Supersymmetry, p-brane duality and hidden space-time dimensions},''
  \href{http://dx.doi.org/10.1103/PhysRevD.54.5203}{{\em Phys. Rev. D} {\bf 54}
  (1996)  5203--5210}, \href{http://arxiv.org/abs/hep-th/9604139}{{\tt
  arXiv:hep-th/9604139}}.

\bibitem{Bars:1996cm}
I.~Bars, ``{Black hole entropy reveals a twelfth dimension},''
  \href{http://dx.doi.org/10.1103/PhysRevD.55.3633}{{\em Phys. Rev. D} {\bf 55}
  (1997)  3633--3641}, \href{http://arxiv.org/abs/hep-th/9610074}{{\tt
  arXiv:hep-th/9610074}}.

\bibitem{Hewson:1996yh}
S.~Hewson and M.~Perry, ``{The Twelve-dimensional super (2+2)-brane},''
  \href{http://dx.doi.org/10.1016/S0550-3213(97)00120-X}{{\em Nucl. Phys. B}
  {\bf 492} (1997)  249--277}, \href{http://arxiv.org/abs/hep-th/9612008}{{\tt
  arXiv:hep-th/9612008}}.

\bibitem{Kutasov:1996zm}
D.~Kutasov, E.~J. Martinec, and M.~O'Loughlin, ``{Vacua of M-theory and
  $\mathcal{N}=2$ Strings},''
  \href{http://dx.doi.org/10.1016/0550-3213(96)00303-3}{{\em Nucl. Phys. B}
  {\bf 477} (1996)  675--700}, \href{http://arxiv.org/abs/hep-th/9603116}{{\tt
  arXiv:hep-th/9603116}}.

\bibitem{Kutasov:1996vh}
D.~Kutasov and E.~J. Martinec, ``{M-branes and $\mathcal{N}=2$ Strings},''
  \href{http://dx.doi.org/10.1088/0264-9381/14/9/008}{{\em Class. Quant. Grav.}
  {\bf 14} (1997)  2483--2516}, \href{http://arxiv.org/abs/hep-th/9612102}{{\tt
  arXiv:hep-th/9612102}}.

\bibitem{Tseytlin:1996ne}
A.~A. Tseytlin, ``{Type IIB instanton as a wave in twelve-dimensions},''
  \href{http://dx.doi.org/10.1103/PhysRevLett.78.1864}{{\em Phys. Rev. Lett.}
  {\bf 78} (1997)  1864--1867}, \href{http://arxiv.org/abs/hep-th/9612164}{{\tt
  arXiv:hep-th/9612164}}.

\bibitem{Bars:1997bz}
I.~Bars and C.~Kounnas, ``{Theories with two times},''
  \href{http://dx.doi.org/10.1016/S0370-2693(97)00452-8}{{\em Phys. Lett. B}
  {\bf 402} (1997)  25--32}, \href{http://arxiv.org/abs/hep-th/9703060}{{\tt
  arXiv:hep-th/9703060}}.

\bibitem{Bars:1997xb}
I.~Bars and C.~Kounnas, ``{String and particle with two times},''
  \href{http://dx.doi.org/10.1103/PhysRevD.56.3664}{{\em Phys. Rev. D} {\bf 56}
  (1997)  3664--3671}, \href{http://arxiv.org/abs/hep-th/9705205}{{\tt
  arXiv:hep-th/9705205}}.

\bibitem{Nishino:1997gq}
H.~Nishino, ``{Supergravity in (10 + 2)-dimensions as consistent background for
  superstring},'' \href{http://dx.doi.org/10.1016/S0370-2693(98)00374-8}{{\em
  Phys. Lett. B} {\bf 428} (1998)  85--94},
  \href{http://arxiv.org/abs/hep-th/9703214}{{\tt arXiv:hep-th/9703214}}.

\bibitem{Nishino:1997sw}
H.~Nishino, ``{$\mathcal{N}=2$ Chiral Supergravity in (10+2)-dimensions as
  Consistent Background for Super(2+2)-brane},''
  \href{http://dx.doi.org/10.1016/S0370-2693(98)00924-1}{{\em Phys. Lett. B}
  {\bf 437} (1998)  303--314}, \href{http://arxiv.org/abs/hep-th/9706148}{{\tt
  arXiv:hep-th/9706148}}.

\bibitem{Hewson:1997wv}
S.~F. Hewson, ``{An Approach to F theory},''
  \href{http://dx.doi.org/10.1016/S0550-3213(98)00521-5}{{\em Nucl. Phys. B}
  {\bf 534} (1998)  513--530}, \href{http://arxiv.org/abs/hep-th/9712017}{{\tt
  arXiv:hep-th/9712017}}.

\bibitem{Linch:2015lwa}
W.~D. Linch and W.~Siegel, ``{F-theory Superspace},''
  \href{http://arxiv.org/abs/1501.02761}{{\tt arXiv:1501.02761 [hep-th]}}.

\bibitem{Linch:2015fya}
W.~D. Linch, III and W.~Siegel, ``{F-theory from Fundamental Five-branes},''
  \href{http://dx.doi.org/10.1007/JHEP02(2021)047}{{\em JHEP} {\bf 02} (2021)
  047}, \href{http://arxiv.org/abs/1502.00510}{{\tt arXiv:1502.00510
  [hep-th]}}.

\bibitem{Heckman:2017uxe}
J.~J. Heckman and L.~Tizzano, ``{6D Fractional Quantum Hall Effect},''
  \href{http://dx.doi.org/10.1007/JHEP05(2018)120}{{\em JHEP} {\bf 05} (2018)
  120}, \href{http://arxiv.org/abs/1708.02250}{{\tt arXiv:1708.02250
  [hep-th]}}.

\bibitem{Heckman:2018mxl}
J.~J. Heckman, C.~Lawrie, L.~Lin, and G.~Zoccarato, ``{F-theory and Dark
  Energy},'' \href{http://dx.doi.org/10.1002/prop.201900057}{{\em Fortsch.
  Phys.} {\bf 67} (2019) no.~10, 1900057},
  \href{http://arxiv.org/abs/1811.01959}{{\tt arXiv:1811.01959 [hep-th]}}.

\bibitem{Heckman:2019dsj}
J.~J. Heckman, C.~Lawrie, L.~Lin, J.~Sakstein, and G.~Zoccarato, ``{Pixelated
  Dark Energy},'' \href{http://dx.doi.org/10.1002/prop.201900071}{{\em Fortsch.
  Phys.} {\bf 67} (2019) no.~11, 1900071},
  \href{http://arxiv.org/abs/1901.10489}{{\tt arXiv:1901.10489 [hep-th]}}.

\bibitem{Dvali:1999hn}
G.~R. Dvali, G.~Gabadadze, and G.~Senjanovic, ``{Constraints on Extra Time
  Dimensions},'' \href{http://arxiv.org/abs/hep-ph/9910207}{{\tt
  arXiv:hep-ph/9910207}}.

\bibitem{Casalbuoni:1975hx}
R.~Casalbuoni, ``{Relativity and Supersymmetries},''
  \href{http://dx.doi.org/10.1016/0370-2693(76)90044-7}{{\em Phys. Lett. B}
  {\bf 62} (1976)  49--50}.

\bibitem{Casalbuoni:1975bj}
R.~Casalbuoni, ``{On the Quantization of Systems with Anticommutating
  Variables},'' \href{http://dx.doi.org/10.1007/BF02748689}{{\em Nuovo Cim. A}
  {\bf 33} (1976)  115}.

\bibitem{Casalbuoni:1976tz}
R.~Casalbuoni, ``{The Classical Mechanics for Bose-Fermi Systems},''
  \href{http://dx.doi.org/10.1007/BF02729860}{{\em Nuovo Cim. A} {\bf 33}
  (1976)  389}.

\bibitem{Schwarz:1982pf}
J.~H. Schwarz and P.~Van~Nieuwenhuizen, ``{SPECULATIONS CONCERNING A FERMIONIC
  SUBSTRUCTURE OF SPACE-TIME},''
  \href{http://dx.doi.org/10.1007/BF02817145}{{\em Lett. Nuovo Cim.} {\bf 34}
  (1982)  21--25}.

\bibitem{Ferrara:2000mm}
S.~Ferrara and M.~A. Lledo, ``{Some Aspects of Deformations of Supersymmetric
  Field Theories},''
  \href{http://dx.doi.org/10.1088/1126-6708/2000/05/008}{{\em JHEP} {\bf 05}
  (2000)  008}, \href{http://arxiv.org/abs/hep-th/0002084}{{\tt
  arXiv:hep-th/0002084}}.

\bibitem{Klemm:2001yu}
D.~Klemm, S.~Penati, and L.~Tamassia, ``{Non(anti)commutative Superspace},''
  \href{http://dx.doi.org/10.1088/0264-9381/20/13/333}{{\em Class. Quant.
  Grav.} {\bf 20} (2003)  2905--2916},
  \href{http://arxiv.org/abs/hep-th/0104190}{{\tt arXiv:hep-th/0104190}}.

\bibitem{Abbaspur:2002xj}
R.~Abbaspur, ``{Generalized Noncommutative Supersymmetry from a New Gauge
  Symmetry},'' \href{http://arxiv.org/abs/hep-th/0206170}{{\tt
  arXiv:hep-th/0206170}}.

\bibitem{deBoer:2003dpn}
J.~de~Boer, P.~A. Grassi, and P.~van Nieuwenhuizen, ``{Non-commutative
  superspace from string theory},''
  \href{http://dx.doi.org/10.1016/j.physletb.2003.08.071}{{\em Phys. Lett. B}
  {\bf 574} (2003)  98--104}, \href{http://arxiv.org/abs/hep-th/0302078}{{\tt
  arXiv:hep-th/0302078}}.

\bibitem{Ooguri:2003qp}
H.~Ooguri and C.~Vafa, ``{The $C$-Deformation of Gluino and Non-Planar
  Diagrams},'' \href{http://dx.doi.org/10.4310/ATMP.2003.v7.n1.a3}{{\em Adv.
  Theor. Math. Phys.} {\bf 7} (2003) no.~1, 53--85},
\href{http://arxiv.org/abs/hep-th/0302109}{{\tt arXiv:hep-th/0302109
  [hep-th]}}.
%%CITATION = HEP-TH/0302109;%%.

\bibitem{Ooguri:2003tt}
H.~Ooguri and C.~Vafa, ``{Gravity Induced $C$-Deformation},''
  \href{http://dx.doi.org/10.4310/ATMP.2003.v7.n3.a2}{{\em Adv. Theor. Math.
  Phys.} {\bf 7} (2003) no.~3, 405--417},
\href{http://arxiv.org/abs/hep-th/0303063}{{\tt arXiv:hep-th/0303063
  [hep-th]}}.
%%CITATION = HEP-TH/0303063;%%.

\bibitem{Seiberg:2003yz}
N.~Seiberg, ``{Noncommutative Superspace, $\mathcal{N} = \frac{1}{2}$
  Supersymmetry, Field Theory and String Theory},''
  \href{http://dx.doi.org/10.1088/1126-6708/2003/06/010}{{\em JHEP} {\bf 06}
  (2003)  010},
\href{http://arxiv.org/abs/hep-th/0305248}{{\tt arXiv:hep-th/0305248
  [hep-th]}}.
%%CITATION = HEP-TH/0305248;%%.

\bibitem{Britto:2003aj}
R.~Britto, B.~Feng, and S.-J. Rey, ``{Deformed superspace, $\mathcal{N} = 1/2$
  supersymmetry and nonrenormalization theorems},''
  \href{http://dx.doi.org/10.1088/1126-6708/2003/07/067}{{\em JHEP} {\bf 07}
  (2003)  067}, \href{http://arxiv.org/abs/hep-th/0306215}{{\tt
  arXiv:hep-th/0306215}}.

\bibitem{Berkovits:2003kj}
N.~Berkovits and N.~Seiberg, ``{Superstrings in Graviphoton Background and
  $\mathcal{N}=1/2 + 3/2$ Supersymmetry},''
  \href{http://dx.doi.org/10.1088/1126-6708/2003/07/010}{{\em JHEP} {\bf 07}
  (2003)  010},
\href{http://arxiv.org/abs/hep-th/0306226}{{\tt arXiv:hep-th/0306226
  [hep-th]}}.
%%CITATION = HEP-TH/0306226;%%.

\bibitem{Cortes:2019mfa}
V.~Cortés, L.~Gall, and T.~Mohaupt, ``{Four-dimensional vector multiplets in
  arbitrary signature},''
\href{http://arxiv.org/abs/1907.12067}{{\tt arXiv:1907.12067 [hep-th]}}.
%%CITATION = ARXIV:1907.12067;%%.

\bibitem{Cortes:2020shr}
V.~Cort\'es, L.~Gall, and T.~Mohaupt, ``{Four-dimensional vector multiplets in
  arbitrary signature (II)},''
  \href{http://dx.doi.org/10.1142/S0219887820501510}{{\em Int. J. Geom. Meth.
  Mod. Phys.} {\bf 17} (2020) no.~10, 2050151}.

\bibitem{Gall:2021tiu}
L.~Gall and T.~Mohaupt, ``{Supersymmetry algebras in arbitrary signature and
  their R-symmetry groups},''
  \href{http://dx.doi.org/10.1007/JHEP10(2021)203}{{\em JHEP} {\bf 10} (2021)
  203}, \href{http://arxiv.org/abs/2108.05109}{{\tt arXiv:2108.05109
  [hep-th]}}.

\bibitem{Witten:1994cga}
E.~Witten, ``{Is supersymmetry really broken?},''
  \href{http://dx.doi.org/10.1142/S0217751X95000590}{{\em Int. J. Mod. Phys. A}
  {\bf 10} (1995)  1247--1248}, \href{http://arxiv.org/abs/hep-th/9409111}{{\tt
  arXiv:hep-th/9409111}}.

\bibitem{Kontsevich:2021dmb}
M.~Kontsevich and G.~Segal, ``{Wick Rotation and the Positivity of Energy in
  Quantum Field Theory},'' \href{http://dx.doi.org/10.1093/qmath/haab027}{{\em
  Quart. J. Math. Oxford Ser.} {\bf 72} (2021) no.~1-2, 673--699},
  \href{http://arxiv.org/abs/2105.10161}{{\tt arXiv:2105.10161 [hep-th]}}.

\bibitem{Witten:2021nzp}
E.~Witten, ``{A Note On Complex Spacetime Metrics},''
  \href{http://arxiv.org/abs/2111.06514}{{\tt arXiv:2111.06514 [hep-th]}}.

\bibitem{Srednyak:2013ylj}
S.~Srednyak and G.~Sterman, ``{Perturbation theory in (2,2) signature},''
  \href{http://dx.doi.org/10.1103/PhysRevD.87.105017}{{\em Phys. Rev. D} {\bf
  87} (2013) no.~10, 105017}, \href{http://arxiv.org/abs/1302.4290}{{\tt
  arXiv:1302.4290 [hep-th]}}.

\bibitem{Carroll:2003st}
S.~M. Carroll, M.~Hoffman, and M.~Trodden, ``{Can the dark energy
  equation-of-state parameter $w$ be less than $−1$?},''
  \href{http://dx.doi.org/10.1103/PhysRevD.68.023509}{{\em Phys. Rev. D} {\bf
  68} (2003)  023509}, \href{http://arxiv.org/abs/astro-ph/0301273}{{\tt
  arXiv:astro-ph/0301273}}.

\bibitem{Cline:2003gs}
J.~M. Cline, S.~Jeon, and G.~D. Moore, ``{The Phantom menaced: Constraints on
  low-energy effective ghosts},''
  \href{http://dx.doi.org/10.1103/PhysRevD.70.043543}{{\em Phys. Rev. D} {\bf
  70} (2004)  043543}, \href{http://arxiv.org/abs/hep-ph/0311312}{{\tt
  arXiv:hep-ph/0311312}}.

\bibitem{Wetterich:2010ni}
C.~Wetterich, ``{Spinors in euclidean field theory, complex structures and
  discrete symmetries},''
  \href{http://dx.doi.org/10.1016/j.nuclphysb.2011.06.013}{{\em Nucl. Phys. B}
  {\bf 852} (2011)  174--234}, \href{http://arxiv.org/abs/1002.3556}{{\tt
  arXiv:1002.3556 [hep-th]}}.

\bibitem{vanNieuwenhuizen:1996tv}
P.~van Nieuwenhuizen and A.~Waldron, ``{On Euclidean spinors and Wick
  rotations},'' \href{http://dx.doi.org/10.1016/S0370-2693(96)01251-8}{{\em
  Phys. Lett. B} {\bf 389} (1996)  29--36},
  \href{http://arxiv.org/abs/hep-th/9608174}{{\tt arXiv:hep-th/9608174}}.

\bibitem{Wess:1992cp}
J.~Wess and J.~Bagger, {\em {Supersymmetry and Supergravity}}.
\newblock Princeton University Press, Princeton, NJ, USA,
1992.
\newblock
%%CITATION = INSPIRE-350988;%%.

\bibitem{DeWitt:2012mdz}
B.~S. DeWitt, \href{http://dx.doi.org/10.1017/CBO9780511564000}{{\em
  {Supermanifolds}}}.
\newblock Cambridge Monographs on Mathematical Physics. Cambridge Univ. Press,
  Cambridge, UK, 5, 2012.

\bibitem{Gates:1983nr}
S.~J. Gates, M.~T. Grisaru, M.~Rocek, and W.~Siegel, {\em {Superspace Or One
  Thousand and One Lessons in Supersymmetry}}, vol.~58 of {\em Frontiers in
  Physics}.
\newblock Benjamin/Cummings, 1983.
\newblock \href{http://arxiv.org/abs/hep-th/0108200}{{\tt
  arXiv:hep-th/0108200}}.

\bibitem{Kawai:2003yf}
H.~Kawai, T.~Kuroki, and T.~Morita, ``{Dijkgraaf-Vafa theory as large-$N$
  reduction},'' \href{http://dx.doi.org/10.1016/S0550-3213(03)00408-5}{{\em
  Nucl. Phys. B} {\bf 664} (2003)  185--212},
  \href{http://arxiv.org/abs/hep-th/0303210}{{\tt arXiv:hep-th/0303210}}.

\bibitem{Chepelev:2003ga}
I.~Chepelev and C.~Ciocarlie, ``{A Path integral approach to noncommutative
  superspace},'' \href{http://dx.doi.org/10.1088/1126-6708/2003/06/031}{{\em
  JHEP} {\bf 06} (2003)  031}, \href{http://arxiv.org/abs/hep-th/0304118}{{\tt
  arXiv:hep-th/0304118}}.

\bibitem{David:2003ke}
J.~R. David, E.~Gava, and K.~S. Narain, ``{Konishi anomaly approach to
  gravitational $F$-terms},''
  \href{http://dx.doi.org/10.1088/1126-6708/2003/09/043}{{\em JHEP} {\bf 09}
  (2003)  043}, \href{http://arxiv.org/abs/hep-th/0304227}{{\tt
  arXiv:hep-th/0304227}}.

\bibitem{Arkani-Hamed:2001vvu}
N.~Arkani-Hamed, T.~Gregoire, and J.~G. Wacker, ``{Higher Dimensional
  Supersymmetry in 4D Superspace},''
  \href{http://dx.doi.org/10.1088/1126-6708/2002/03/055}{{\em JHEP} {\bf 03}
  (2002)  055}, \href{http://arxiv.org/abs/hep-th/0101233}{{\tt
  arXiv:hep-th/0101233}}.

\bibitem{Dijkgraaf:2003xk}
R.~Dijkgraaf and C.~Vafa, ``{$\mathcal{N}=1$ Supersymmetry, Deconstruction, and
  Bosonic Gauge Theories},'' \href{http://arxiv.org/abs/hep-th/0302011}{{\tt
  arXiv:hep-th/0302011}}.

\bibitem{Terashima:2003ri}
S.~Terashima and J.-T. Yee, ``{Comments on noncommutative superspace},''
  \href{http://dx.doi.org/10.1088/1126-6708/2003/12/053}{{\em JHEP} {\bf 12}
  (2003)  053}, \href{http://arxiv.org/abs/hep-th/0306237}{{\tt
  arXiv:hep-th/0306237}}.

\bibitem{Colladay:1996iz}
D.~Colladay and V.~A. Kostelecky, ``{CPT Violation and the Standard Model},''
  \href{http://dx.doi.org/10.1103/PhysRevD.55.6760}{{\em Phys. Rev. D} {\bf 55}
  (1997)  6760--6774}, \href{http://arxiv.org/abs/hep-ph/9703464}{{\tt
  arXiv:hep-ph/9703464}}.

\end{thebibliography}\endgroup

\end{document}